# Emergence and Collapse of Leader–Follower Flocks through Phantom Targets: Surprise Nullification in Bayesian and Inverse Bayesian Inference

Shuji Shinohara[a,*], Daiki Morita[a], Yoshihiro Nakajima[b], Takeshi Takano[c], Masakazu Higuchi[d], Ung-il Chung[e], and Yukio-Pegio Gunji[f]

[a] *School of Science and Engineering, Tokyo Denki University, Hatoyama, Saitama, Japan*
[b] *Graduate School of Economics, Osaka Metropolitan University, Osaka, Japan*
[c] *Department of Advanced Energy, Graduate School of Frontier Sciences, The University of Tokyo, Kashiwa, Chiba, Japan*
[d] *Department of Fundamental Engineering (Information, Electronics, and Optics Course), Utsunomiya University, Utsunomiya, Tochigi, Japan*
[e] *Department of Bioengineering, Graduate School of Engineering, The University of Tokyo, Tokyo, Japan*
[f] *Department of Intermedia Art and Science, School of Fundamental Science and Engineering, Waseda University, Tokyo, Japan*

* Corresponding author: s.shinohara@mail.dendai.ac.jp

**ABSTRACT**. In Bayesian and inverse Bayesian (BIB) inference, large prediction errors can trigger surprise nullification, which expands belief and likelihood variances by a common factor while retaining the variance-ratio update rule of standard Bayesian inference. This allows belief uncertainty to grow without necessarily increasing the learning rate. We investigated how this property shapes collective motion in a minimal model of three agents. Each agent used BIB inference to predict the positions of the other two and moved toward their predicted positions, without alignment rules, repulsion, or designated leaders. Across 100 simulation trials, the agents repeatedly formed and dissolved leader–follower flocks. Three criteria—front position, heading stability, and initiation of flock formation—identified the same leader in 79% of eligible flocks, and leadership rotated among agents. Leader–follower flocks typically arose when an agent nearly ceased incorporating observations of one partner (a link cut) and continued toward an extrapolated position detached from that partner's actual location (a phantom target). Link cuts generally originated from two consecutive redrawings of the likelihood variance (resets): a very small value first reduced belief variance, and a subsequent ordinary value sharply increased the likelihood-to-belief variance ratio and suppressed learning. While the phantom target persisted, the agent moved nearly straight and the others followed. Flocks typically collapsed as the returning phantom target brought the leader's forward and backward attractions into balance; link recovery followed, rather than triggered, the collapse. Control models revealed complementary roles for the two sides of nullification: without nullification, cuts almost never recovered, whereas expanding only belief variance led to recovery within a few steps. Joint expansion permitted cuts to persist despite growing belief uncertainty and enabled later recovery through resets that reassigned likelihood variance without changing the nullification strength. Thus, surprise nullification supports both the persistence and eventual resolution of inaccurate individual predictions, allowing transient collective order to emerge without prescribed leadership.

## I. INTRODUCTION

In standard Bayesian (Kalman) filtering of a target whose state does not change, repeated observations accumulate information, reducing both belief uncertainty and the degree to which each new observation updates the belief (the gain, or learning rate) [1]. This property is effective in stable environments, but in nonstationary environments the constraint imposed by past beliefs and the reduced sensitivity to new observations (the learning rate) can prevent the belief from tracking changes. Consequently, methods that weaken the influence of beliefs based on past observations through forgetting, as well as mechanisms that adjust the learning rate according to environmental volatility, have been studied [2–5].

Large prediction errors, however, can arise either from a change in the state of the target [6] or from a transient outlier [7]. Immediately increasing the learning rate in response to any large error, including outliers, can cause overreactions to incidental observations; weakening confidence in the existing belief and strongly incorporating new observations therefore should not always be coupled. To address this problem, models have been proposed that learn while distinguishing environmental volatility from observation stochasticity [8]. An alternative is to defer attributing an error to either source while weakening confidence in the existing belief without abruptly revising the prediction, thereby preparing for later updates.

Bayesian and inverse Bayesian (BIB) inference was originally proposed by Gunji et al. [9], and a model incorporating a symmetry bias was subsequently formulated [10, 11]. Within this framework, "nullification" was defined as an operation that weakens confidence in both the existing belief and the prediction with a common strength $\beta$ [12] ($\beta = 0$ corresponds to standard Bayesian inference). In the Gaussian case, nullification increases the belief and likelihood variances, so that their ratio equals that of the standard Bayesian update. Consequently, when the likelihood variance is sufficiently large relative to the belief variance, confidence in the existing belief can be weakened, while the learning rate is kept low. As a result, even when observations and predictions diverge greatly, a state can arise in which, although the cause of the error remains unclear, the prediction is not revised substantially, and only the uncertainty grows. This state can be resolved by redrawing the observation reliability (likelihood variance) (hereafter, a reset) when, as the prediction error decreases, the redrawn likelihood variance explains the current observation better than the carried-over one (i.e., yields a smaller predictive surprise).

In this study, BIB inference with nullification was applied to three moving agents. BIB inference has previously been used to model the foraging behavior of a single agent [13] and its swarming behavior [14, 15]. In our model, each agent observes and predicts the positions of other agents and moves toward these predicted positions. Because all agents follow the same rule, their interactions make each agent's environment endogenously nonstationary.

This setup differs from conventional flocking models based on explicit interaction rules for alignment and, in many models, repulsion and attraction [16–19] (reviewed in [20, 21]). Diverse collective motions have also been reported in models in which agents anticipate the orientations of their neighbors [22] or extrapolate future positions [23], and a coherent and cohesive swarm has been shown to emerge from mutual anticipation without an alignment rule [24]. A predictive alignment model has been shown to produce cohesive flocks without explicit attraction by selecting movement directions based on the alignment and number of neighbors at their own predicted future positions [25]. Furthermore, models applying active inference, grounded in the free-energy principle [26, 27], to groups of agents [28] and models using allocentric representations of the bearings of other individuals [29] have shown that collective motion can arise without explicit alignment rules. Similarly, our model uses attraction toward the predicted positions of the other agents as its only movement rule and introduces neither alignment nor repulsion. In this setting, we examined how predictions that diverge from reality persist under nullification, how they are subsequently resolved, and how this process affects collective motion.

In animal groups, hierarchical leadership, leadership determined by speed differences, and collective decision-making through adjustments of movement direction have been reported [30–32]. Leadership models in which individuals with goal information are designated in advance have also been proposed [33]. Leader and follower roles can also emerge spontaneously from differences in state between individuals that follow the same rules [34]. In contrast, our model assigns all agents the same rules and designates neither a particular leader nor an external goal. Thus, any leader that emerges must arise endogenously from the inferences and interactions among agents.

A link cut is defined as a state in which an agent incorporates little information from observations of a particular other agent, and control models are used to examine the respective roles of nullification and reset in the onset, persistence, and recovery of link cuts. The simulations show that an agent that has cut a link can become a leader by pursuing a predicted position detached from the observed position (a phantom target), that a flock forms as the other agents follow it, and that the flock is maintained through mutual reinforcement between the leader's straight motion and the followers' movements. They further show that nullification reconciles the persistence and recovery of link cuts by lowering absolute precision while preserving relative precision during a cut, thereby contributing to both the maintenance and collapse of flocks. The remainder of this paper is structured as follows. In Sec. II, BIB inference and the flocking model are formulated. Section III presents the analyses, and Sec. IV examines the role and significance of nullification.

## II. METHODS

### A. Overview

The three agents moved at constant speed on an unbounded plane. Each agent has an independent filter for each of the other two agents, which estimates each partner's position from noisy observations and predicts the partner's position at the next time step. The filters implement BIB inference [12]. Each agent moved in the direction of the sum of the attraction vectors toward the predicted positions of the other agents obtained from its filters.

Below, the update equations are summarized for a single link from agent $i$ to agent $j$. The BIB update rules are derived in Appendix A and in previous work [12]. Time is indicated by a superscript to distinguish it from the agent indices.

### B. Observation and prediction

At time $t$, agent $i$ obtains the position $\mathbf{x}_j^t$ of agent $j$ as an observation $\mathbf{o}_{ij}^t$ that includes isotropic Gaussian noise $\boldsymbol{\varepsilon}_{ij}^t$.

$$\mathbf{o}_{ij}^t = \mathbf{x}_j^t + \boldsymbol{\varepsilon}_{ij}^t, \qquad \boldsymbol{\varepsilon}_{ij}^t \sim \mathcal{N}(\mathbf{0}, \sigma^2\mathbf{I}), \quad \sigma = 1 \tag{1}$$

The observation noise is generated independently for each time $t$ and link $ij$.

Agent $i$ represents its estimate of the position of agent $j$ as a two-dimensional vector $\mathbf{m}_{ij}^t$. The predicted position $\widehat{\mathbf{m}}_{ij}^t$ at time $t$ is obtained by constant-velocity extrapolation from the estimated positions at the two preceding time steps.

$$\widehat{\mathbf{m}}_{ij}^t = \mathbf{m}_{ij}^{t-1} + \left(\mathbf{m}_{ij}^{t-1} - \mathbf{m}_{ij}^{t-2}\right) = 2\,\mathbf{m}_{ij}^{t-1} - \mathbf{m}_{ij}^{t-2} \tag{2}$$

Let $\mathbf{e}_{ij}^t = \mathbf{o}_{ij}^t - \widehat{\mathbf{m}}_{ij}^t$ be the prediction-error vector and $\delta_{ij}^t = \| \mathbf{e}_{ij}^t \|$ its magnitude. In the BIB inference, the learning rate $K_{ij}^t$ is determined from the magnitude of the prediction error.

The estimated position is corrected from the predicted position in the direction of the prediction error vector according to the learning rate.

$$\mathbf{m}_{ij}^t = \widehat{\mathbf{m}}_{ij}^t + K_{ij}^t\,\mathbf{e}_{ij}^t \tag{3}$$

### C. Movement

Agent $i$ sums the vectors toward the next-step predicted positions of the other two agents and moves a constant distance $v_0 = 1$ in that direction.

$$\mathbf{h}_i^t = \sum_{j \neq i} \left(\widehat{\mathbf{m}}_{ij}^{t+1} - \mathbf{x}_i^t\right), \qquad \mathbf{x}_i^{t+1} = \mathbf{x}_i^t + v_0 \frac{\mathbf{h}_i^t}{\| \mathbf{h}_i^t \|} \tag{4}$$

Here, $\widehat{\mathbf{m}}_{ij}^{t+1} = 2\mathbf{m}_{ij}^t - \mathbf{m}_{ij}^{t-1}$ is the predicted position for the next time step, obtained by the same extrapolation as in Eq. (2) after the observation at time $t$ is incorporated. If $\mathbf{h}_i^t = \mathbf{0}$, then the agent does not move. This movement rule is equivalent to moving toward the centroid (midpoint) of the two predicted positions. All agents are updated synchronously; they observe the positions at the same time step, update their estimates, and then move simultaneously.

### D. Determination of the learning rate

This section describes how the learning rate is determined by BIB inference. For simplicity, the link index $ij$ is omitted.

#### *1. General form of nullification*

The joint distribution $p(s, o)$ of the state $s$ and the observation $o$ can be decomposed, using the belief distribution $q(s)$, the predictive distribution $p(o)$ and the lift $\Lambda(s, o) = p(s, o)/[q(s)\,p(o)]$, as

$$p(s, o) = q(s)\,p(o)\,\Lambda(s, o) \tag{5}$$

Here $q(s)$ and $p(o)$ represent the current confidence about the state and the expectation about the observation, respectively, whereas $\Lambda(s, o)$ represents the strength of the coupling relative to the case in which the state and the observation are independent.

Nullification is an operation that simultaneously weakens confidence in the belief and in the prediction with a common strength parameter β ($0 \leq \beta \leq 1$); it is expressed by the following nullification kernel:

$$\mathcal{K}_\beta(s, o) = [q(s)\,p(o)]^{1-\beta} \Lambda(s, o) \tag{6}$$

When $\beta = 0$, $\mathcal{K}_\beta$ coincides with the original joint distribution. Normalizing this kernel over the state $s$ with the observation $o$ fixed yields the belief update and normalizing it over the observation $o$ with the state $s$ fixed yields the likelihood update.

The nullification strength $\beta$ does not act on the lift $\Lambda(s, o)$. Nullification can therefore be interpreted as an operation that weakens the marginal probabilistic weights of the state and the observation while retaining their cross-coupling structure.

When the belief and the likelihood are Gaussian, these two normalizations yield the update equations for the learning rate and variances given in Eq. (7) (Appendix A). However, because our model imposes an additional constraint that fixes the center of the likelihood, this preservation of the coupling structure does not hold exactly for the updated joint distribution. This constraint applies only to the center of the likelihood and does not affect the update in Eq. (7) for the learning rate and variances (Appendix A, Section 3).

#### *2. Application to two-dimensional position estimation*

In our model, the position of another agent is estimated as a two-dimensional vector, whereas the learning rate and nullification strength are computed from the magnitude $\delta^t$ of the prediction-error vector.

Specifically, at each time step, the direction of the prediction-error vector is treated as a one-dimensional axis with the predicted position as the origin. On this axis, the belief about the partner's position is represented by a Gaussian distribution with mean 0 and variance $P^t$, and the likelihood of the observation is represented by a Gaussian distribution with variance $R^t$.

The one-dimensional BIB update rule is applied to the magnitude $\delta^t$ of the prediction error to determine the learning rate, which is then used for the two-dimensional position estimate in Eq. (3). Our model is therefore not an update rule rigorously derived from a two-dimensional Gaussian distribution but an adaptive position-estimation model based on the one-dimensional BIB update rule.

### *3. Nullification update and reset update*

Let $(\tilde{R}^t, \tilde{\beta}^t)$ be the candidate likelihood variance and nullification strength generated at the previous time step.

The likelihood variance $R^t$ and the nullification strength $\beta^t$ actually used at time $t$ are chosen from the following two update candidates.

**Nullification update:** $(R^t, \beta^t) = (\tilde{R}^t, \tilde{\beta}^t)$. The candidates generated in the previous time step are adopted. When the candidate nullification strength is zero, this reduces to a standard Bayesian update that carries over the likelihood variance from the previous time step.

**Reset update:** $(R^t, \beta^t) = (2R_0 u^t, \tilde{\beta}^t), u^t \sim U(0,1)$. Only the likelihood variance is redrawn; the candidate nullification strength is used, as in the nullification update. $R_0$ is the reference likelihood variance, and $u^t$ is generated independently for each time step and link. Which update is used is determined by the reset criterion described below (Sec. II D 6). The two updates differ only in the likelihood variance; the reset update corresponds to the revision of the observation reliability mentioned in the Introduction. Below, the use of the reset update is simply called a reset.

In [12], $(R^t, \beta^t) = (R_0, 0)$ was used at a reset. In this study, by contrast, the likelihood variance at a reset is not fixed to the reference value but is redrawn from the uniform distribution on the interval $(0, 2R_0)$, whose mean equals $R_0$, and the nullification strength is not changed by a reset. A reset therefore determines only the likelihood variance, and the nullification strength is determined only by Eqs. (9)–(11).

This is a modeling assumption for resetting the observation reliability at a reset and is not derived from the distribution that generates the observation noise. This setting allows very different likelihood variances to be chosen before and after a reset. The dependence on the shape of this distribution is examined in Appendix E.

### *4. Updating the learning rate and variances*

Using the selected $(R^t, \beta^t)$, the learning rate and variances are updated as follows:

$$\begin{aligned} K^t &= \frac{P^t}{P^t + (1-\beta^t)R^t} = \frac{1}{1 + (1-\beta^t)\, R^t/P^t}, \\ P^{t+1} &= \frac{P^t R^t}{P^t + (1-\beta^t)R^t} = K^t R^t, \\ \tilde{R}^{t+1} &= \frac{R^t(P^t + R^t)}{P^t + (1-\beta^t)R^t} = P^{t+1}\left(1 + \frac{R^t}{P^t}\right). \end{aligned} \tag{7}$$

$P^{t+1}$ and $\tilde{R}^{t+1}$ are created by the update at time $t$ and are used at the next time step. Equation (7) gives $\tilde{R}^{t+1}/P^{t+1} = 1 + R^t/P^t$. In other words, the ratio of the likelihood variance to the belief variance after a nullification update (hereafter, the variance ratio) does not depend on $\beta^t$. The variance ratio also equals the ratio of belief precision (the inverse of variance) to observation precision and thus represents their relative precision. Because nullification expands the belief and likelihood variances by the same factor relative to the standard Bayesian update, it can change the absolute magnitudes of both variances without changing the update rule for the variance ratio. In particular, when $R^t/P^t$ is sufficiently large, the belief variance can be increased while keeping the learning rate low.

### *5. Determination of the nullification strength*

The candidate nullification strength for the next time step is determined by minimizing the predictive surprise after the update [12].

Let $\tilde{p}^{t+1}(\cdot; R, \beta) = \int q^{t+1}(s)\tilde{l}^{t+1}(o \mid s)\, ds$ be the predictive distribution constructed from the updated belief $q^{t+1}(s)$ and the candidate likelihood $\tilde{l}^{t+1}(o \mid s)$.

The predictive surprise for the current prediction error $\delta^t$ is then

$$-\log\tilde{p}^{t+1}(\delta^t; R, \beta) = \frac{1}{2}\log\frac{2\pi R\,(2P + R)}{P + (1-\beta)R} + \frac{(1-\beta)^2 R\, \delta^2}{2\,(2P + R)\{P + (1-\beta)R\}} \tag{8}$$

where we write $P = P^t$ and $\delta = \delta^t$ on the right-hand side.

Under the likelihood variance $R^t$ used at time $t$, the nullification strength that minimized this surprise is obtained and used as a candidate for the next time step.

$$\tilde{\beta}^{t+1} = \underset{0 \le \beta \le 1}{arg\,min}[-log\tilde{p}^{t+1}(\delta^t; R^t, \beta)] \tag{9}$$

This optimization problem can be solved in closed form.

$$\tilde{\beta}^{t+1} = \begin{cases} 0, & \delta^2 \le P + R, \\ 1 - x^*, & \delta^2 > P + R, \end{cases} \quad (10)$$

$$x^* = \frac{R\,(2P + R) - 2P\delta^2 + \sqrt{R^2(2P + R)^2 + 4P^2\delta^4}}{2R\,\delta^2} \quad (11)$$

Thus the candidate nullification strength is positive only when $(\delta^t)^2 > P^t + R^t$ ; nullification is determined not by the absolute magnitude of the prediction error but by its magnitude relative to the variance of the predictive distribution.

Note that the nullification strength determined from the prediction error obtained at time $t$ is not applied directly to the learning rate at that time but is used as the candidate for the next time step.

TABLE I. Components of BIB and the control models

| Model | Next-step candidate likelihood variance | Candidate nullification strength | Reset |
|---|---|---|---|
| BIB | Eq. (7) | Automatic | Yes |
| fixed-β BIB | Eq. (7) | $\beta_0 = 0.5$ | Yes |
| random-R BIB | $2R_0u'$ | Automatic | Yes |
| FB | Fixed at $R_0$ | $\beta_0 = 0.05$ | No |
| zero-β BIB | Eq. (7) | $\beta_0 = 0$ | Yes |
| fixed-R BIB | Kept from the previous step | Automatic | Yes |
| Standard Bayes† | Fixed at $R_0$ | $\beta_0 = 0$ | No |

† Shown for reference (not simulated). Because standard Bayes performs neither nullification nor resets, the variance ratio increases monotonically according to $R_0/P^t = R_0/P^0 + t$, and the learning rate approaches zero ($P^0$ is the initial belief variance).

***6. Reset criterion***

The two updates are compared by how well they explain the current prediction error before the update, that is, by the predictive surprise under the pre-update predictive distribution, $S^t(R) = -\log N(\delta^t; 0, P^t + R)$, with $R = 2R_0u^t$ or $R = \tilde{R}^t$. A reset is used if its predictive surprise is less than or equal to that of the nullification update.

$$R^t = \begin{cases} 2R_0u^t, & S^t(2R_0u^t) \le S^t(\tilde{R}^t), \\ \tilde{R}^t, & \text{otherwise.} \end{cases} \quad (12)$$

Because both updates use the same nullification strength, Eq. (12) compares the two likelihood variances only. Since $-\log N(\delta; 0, V) = ½\log(2\pi V) + \delta^2/(2V)$ is smallest at $V = \delta^2$ , a smaller likelihood variance is preferred when the prediction error is small relative to the predictive variance and a larger one when it is large. Unlike the updated predictive distribution in Eq. (8), whose density diverges as $R \to 0$, the pre-update predictive density remains bounded because its variance is at least $P^t$; the criterion therefore does not by itself favor very small likelihood variances (Appendix E 3). In summary, at each time step, Eq. (12) selects the likelihood variance (reset or nullification update), Eq. (7) updates the learning rate and variances, and Eqs. (10) and (11) provide the candidate nullification strength for the next time step.

**E. Control models**

To examine the roles of the components of BIB, we prepared five control models that altered the automatic determination of the nullification strength, the update of the likelihood variance, the presence of resets, or nullification itself (Table I). The observation noise, movement rule, and random seeds were the same as those in the BIB. In fixed-β BIB, the candidate nullification strength $\tilde{\beta}^{t+1}$ given by Eqs. (10) and (11) is replaced with a fixed value $\beta_0 = 0.5$; in random-R BIB, the candidate likelihood variance $\tilde{R}^{t+1}$ given by Eq. (7) is replaced with a uniform random number $2R_0u'$ (which amounts to replacing the likelihood side of nullification with random numbers). Forgetting Bayes (FB) fixes the likelihood variance at $R_0$ , performs no resets, and uses a fixed $\beta_0 = 0.05$ as a forgetting rate; with $\beta_0 = 0$, FB coincides with standard Bayes (Table I). In zero-β BIB, the nullification strength is always zero (fixed-β BIB with $\beta_0 = 0$); this model differs from standard Bayes only in having resets. In fixed-R BIB, only the update of the likelihood variance is removed (which amounts to letting nullification act only on the belief variance), thus the likelihood variance drawn at a reset is maintained until the next reset. All models with resets use the same reset rule [Eq. (12)]. Appendix D presents the formulation of each model.

**F. Simulation settings**

One hundred trials were conducted for each model. Each trial used a different random seed, and the same seed was used across models for the same trial number. The first $10^5$ steps of each trial were discarded as a transient, and the subsequent $10^5$ steps were analyzed. The reference likelihood variance $R_0$ was set to the true observation noise variance $\sigma^2 = 1$. The simulation conditions are listed in Table II.

TABLE II. Simulation settings

| Item | Value |
|---|---|
| Number of agents $N$ | 3 |
| Space | Unbounded two-dimensional plane |
| Observability | Each agent always observes the other two |
| Speed $v_0$ | 1 (per step) |
| Observation noise variance $\sigma^2$ | 1 (per coordinate) |
| Reference likelihood variance $R_0$ | 1 |
| Initial belief variance | 100 |
| Initial positions | Uniform on $[0,100]^2$ |
| Transient | $10^5$ steps |
| Analysis period | $10^5$ steps |
| Number of trials | 100 trials per model |
| Random seeds | Set by trial number; common across models |

## G. Detection of flocks and leaders

### *1. Flock formation and collapse*

The group alignment $A$ is the average of $\cos(\theta_i - \theta_j)$ over all pairs of agents, where $\theta_i$ denotes the headings. For three agents, the alignment lies in the range $-0.5 \leq A \leq 1$, and $A = 1$ when all agents have the same heading.

The flock formation time $t_F$ is defined as the first time step of an interval in which $A \geq 0.90$ holds for 30 consecutive steps. The flock collapse time is the first time step after $t_F + 30$ at which $A < 0.5$. The search for the next flock begins immediately after the collapse of the previous flock.

### *2. Detection of link cuts*

Link cuts are detected for each link from the ratio $R/P$ of the likelihood variance $R$ actually used at that time step to the belief variance $P$. We call a state satisfying $R/P > T$ ($T = 1{,}000$) a "link cut" and define the time at which $R/P \leq T$ again as the recovery of the link. The threshold $T = 1{,}000$ is used only for the analysis of the results and affects neither the motion nor the inference updates of the agents. Cuts already in progress at the start of the analysis period are excluded because their onset is not observed. A cut whose recovery also falls within the analysis period is called a completed cut; the numbers and durations of cuts (Table III) refer to completed cuts, whereas the analyses of the variance ratio at the onset of a cut (Figs. 5b and 8) also include the cuts that had not recovered by the end of the analysis period.

The learning rate is $K = \frac{1}{1+(1-\beta)R/P}$. Hence the condition $R/P > T$ alone does not uniquely determine the learning rate. In our simulations, however, the nullification strength $\beta$ during cuts is small at most time steps, and the learning rate is generally below $10^{-3}$.

### *3. Leader identification*

Flock leaders are identified using the following three criteria.

**Formation initiator:** the agent whose low-turning interval, continuing without interruption up to the flock formation time $t_F$, began earliest. A low-turning interval is a sequence of consecutive steps in which the turning angle at every step satisfies $|\Delta\theta| < 0.10$ rad. This interval is searched back to 200 steps before $t_F$, but not beyond the collapse of the previous flock.

**Stable agent:** the agent with the smallest mean turning angle per step, $\overline{|\Delta\theta|}$, during the flock.

**Front agent:** the positions of the agents are projected onto the mean heading of the flock, and the agent with the largest projection for the greatest number of steps during the flock is designated as the front agent.

If a criterion results in a tie, that criterion is treated as undetermined. An agent is identified as the leader when all three criteria point to it. The analyses of the leader's cut links in Secs. III C–III E focus on the cut links that this leader holds at flock formation.

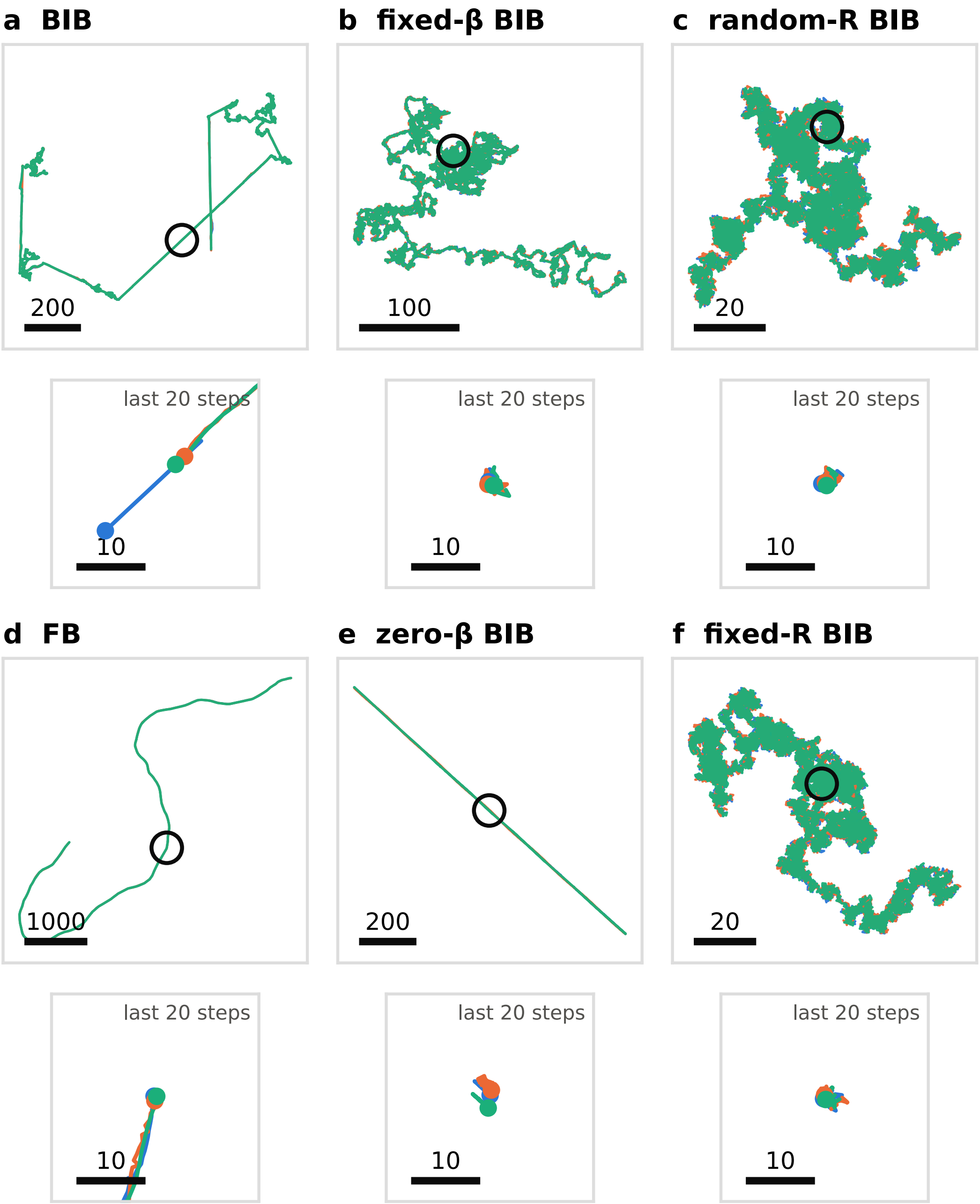

**FIG. 1.** Trajectories of the six models ($N = 3$). Top: 10,000 steps of one trial. Bottom: the three agents at the circled step, with their last 20 steps (equal scale). (a) Bayesian and inverse Bayesian (BIB), (b) fixed-β BIB ($\beta_0 = 0.5$), (c) random-R BIB, (d) Forgetting Bayes (FB) ($\beta_0 = 0.05$), (e) zero-β BIB, (f) fixed-R BIB. In (e) the agents stay within a few units of one another and reverse their headings almost every step, while the group drifts along a straight line. Note the different scale bars.

## III. RESULTS

### A. Collective motion of the six models

Figure 1 shows the trajectories of the $10^4$ steps from the analysis period of one trial for each model. In BIB, the three agents alternated between long, nearly straight segments with sharp turns and moved over a region more than 1,000 units across $10^4$ steps (Fig. 1a, top). During straight segments, one agent was in front and the other two followed in single file behind it (Fig. 1a, bottom). FB also moved over a wide region, but its trajectories curved smoothly and the three agents translated side by side (Fig. 1d). In contrast, in fixed-β BIB, random-R BIB, and fixed-R BIB, the three agents circled around one another and remained in a small region (Fig. 1b, c, f). In zero-β BIB, the three agents stayed within a few units of one another but reversed their headings almost every step, while the group drifted along a straight line (Fig. 1e); in other trials, they remained several hundred units apart (Sec. III G). Thus, straight motion in single file occurred only in BIB.

### B. Emergence of leaders

Of the 1,603 BIB flocks observed across 100 trials, 1,411 flocks were analyzed for which the formation initiator could be determined, the onset of its low-turning interval lay within the search window, and the flock collapsed before the end of the record (18,197 flocks for FB). In BIB, the fraction of flocks in which two of the three criteria named the same agent was 0.83–0.91 for every pair, and the fraction in which all three agreed was 0.79, far above the chance levels obtained by permuting the agents named by each criterion across flocks (about 1/3 for a pair and 1/9 for all three) (Fig. 2a). In FB, by contrast, agreement was 0.16–0.40 for pairs and 0.07 for all three, close to chance. Thus, most BIB flocks had a single identifiable leader: an agent that occupied the front position, maintained a stable heading, and initiated flock formation.

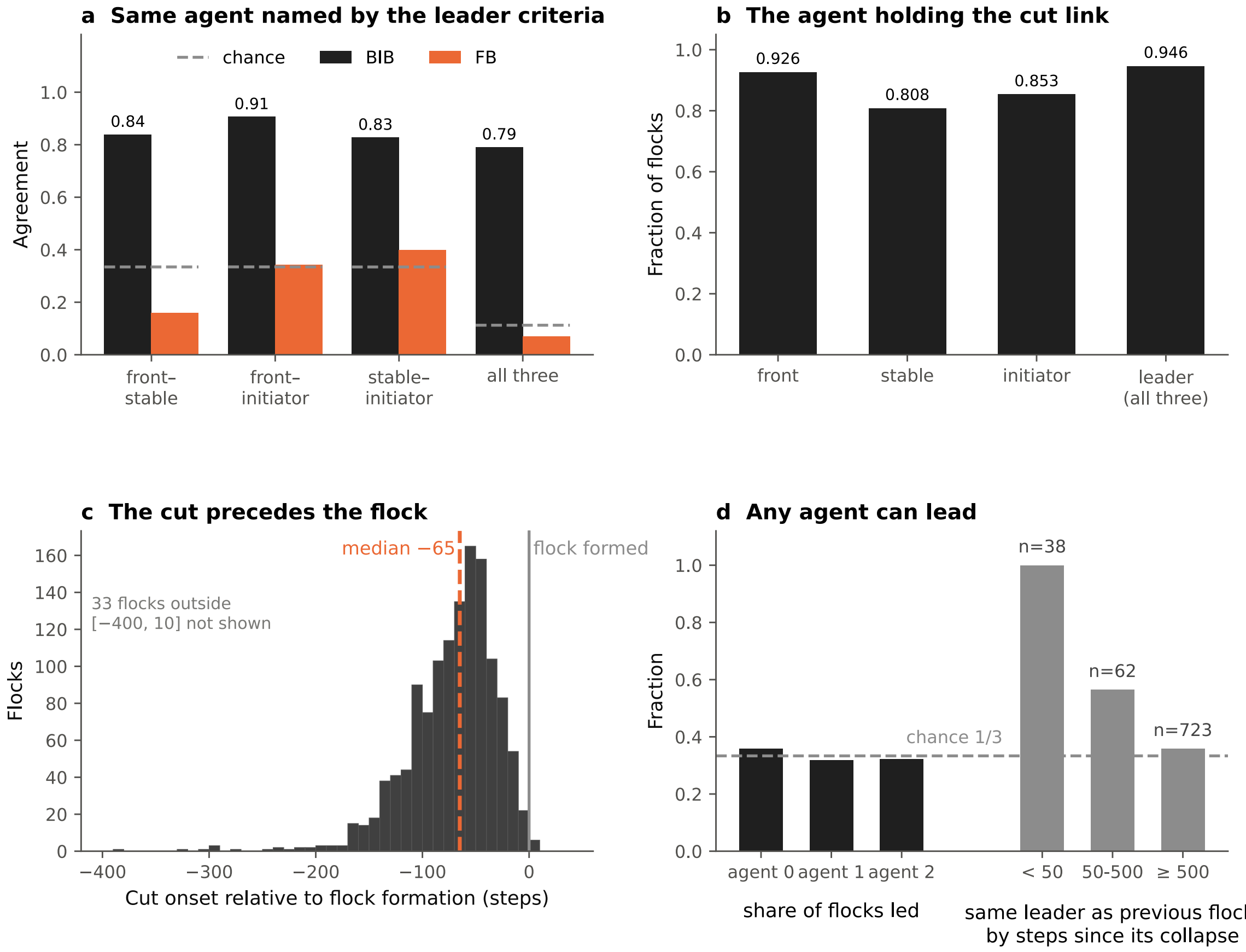

**FIG. 2.** Leaders of BIB flocks (100 trials; 1,603 flocks). (a–c) The 1,411 flocks for which the formation initiator was determined, the onset of its low-turning interval fell within the search window, and the flock collapsed before the end of the record (FB: 18,197 flocks). (a) Agreement between the three leader criteria (front agent, stable agent, formation initiator) in BIB and FB; dashed, chance (agents permuted across flocks: about 1/3 for a pair, 1/9 for all three). (b) Fraction of flocks in which the agent named by each criterion is the agent holding the cut link (the agent whose cut links cover at least half of the flock). (c) Onset of that agent's first cut link relative to flock formation (median −65 steps; 33 flocks outside the plotted range are not shown). (d) Left: share of the 1,175 flocks with a leader (the three criteria agree; out of the 1,483 flocks with a determined initiator, without the conditions used in panels a–c) that are led by each agent. Right: probability that a flock has the same leader as the previous flock of the same trial, as a function of the number of steps between the collapse of the previous flock and the formation of this one (consecutive flocks with a leader; n above the bars). Dashed line, chance 1/3.

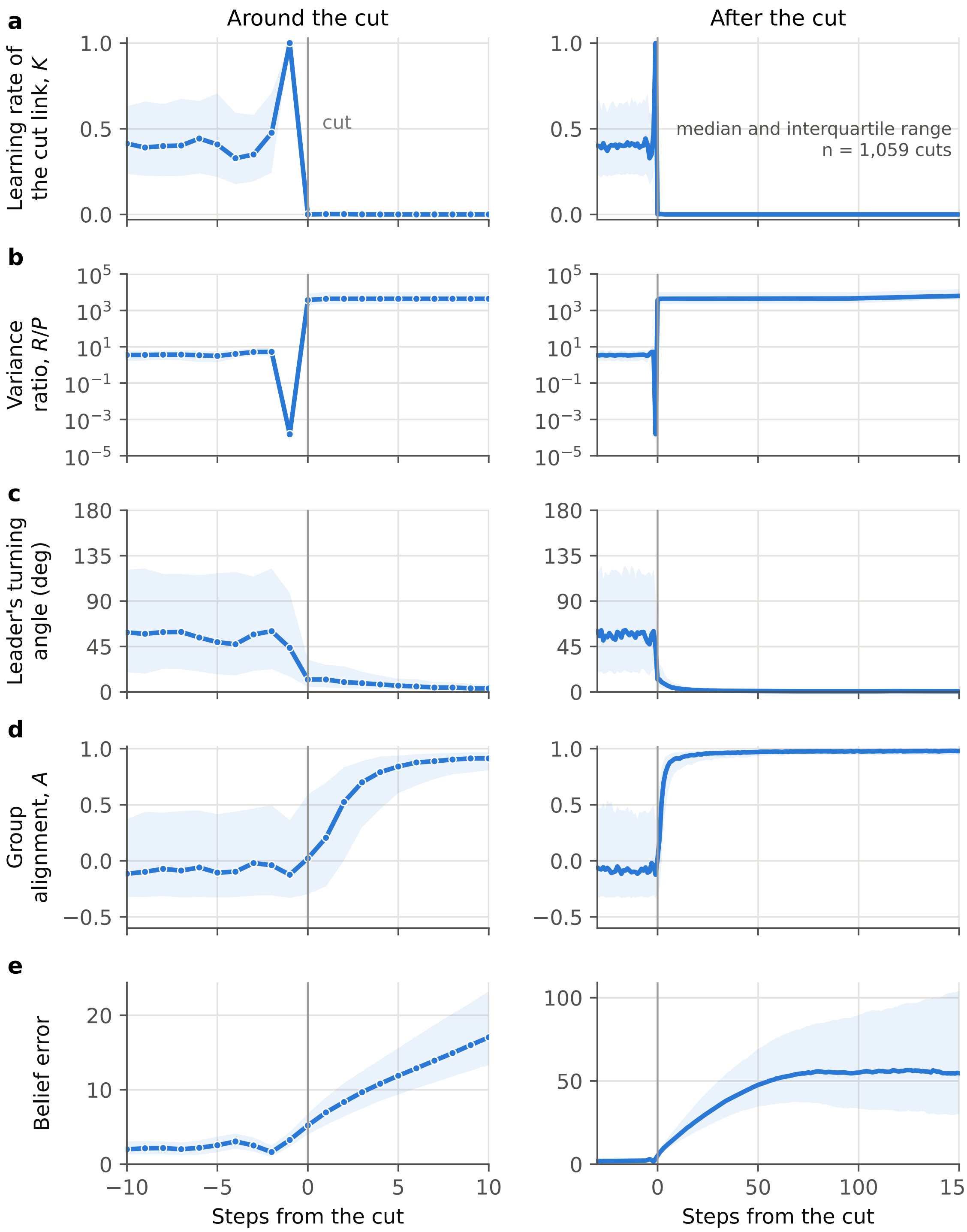

**FIG. 3.** The leader's cut link aligned at the cut (leaders named by all three criteria; n = 1,059 cuts; median and interquartile range). Left, ±10 steps; right, −30 to 150 steps. (a) Learning rate K of the cut link. (b) R/P. (c) Leader's turning angle. (d) Group alignment A. (e) Belief error. One step before the cut, a reset selected at a small prediction error draws a small R (median $3.4\times10^{-4}$): $K \approx 1$ and P falls to R. At the cut a second reset with an ordinary R gives $R/P \approx R(0)/R(-1) \approx 3.7\times10^{3}$ (median) and $K \approx 0$; R/P then stays nearly constant, increasing by only 1 per step [Eq. (7)]. This holds for 99% of the 1,059 cuts.

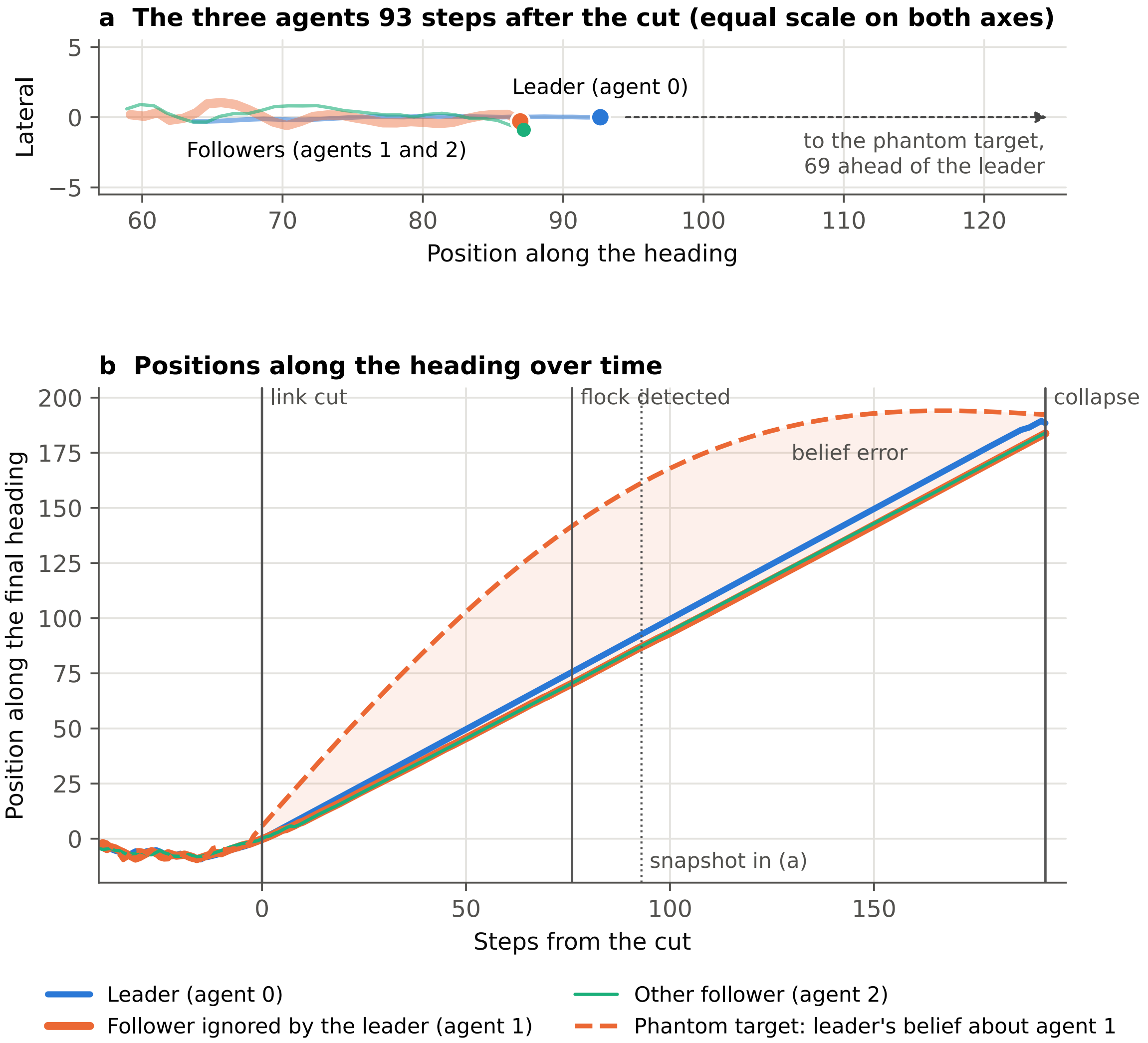

**FIG. 4.** A phantom target (one cut). (a) The three agents 93 steps after the cut; the leader (agent 0) heads toward the phantom target 69 units ahead. (b) Positions along the final heading: the leader, the follower whose link the leader cut (agent 1), the other follower (agent 2), and the phantom target, i.e. the leader's belief about agent 1. Shading indicates belief error.

The agent that held a cut link for at least half of the flock's duration coincided with the leader identified using all three criteria in 95% of flocks (Fig. 2b). Its link cut began before flock formation in almost all flocks (99.1%), with a median lead of 65 steps (Fig.

2c). Thus, the leader was the agent that had cut its link to another agent before the flock formed.

Leadership was not restricted to a particular agent. For this analysis, all 1,483 flocks for which the formation initiator could be determined were used, without the window and collapse conditions applied above, because these conditions exclude many flocks that re-formed shortly after the previous one collapsed. In the 1,175 of these flocks for which the three criteria agreed (79%), each of the three agents led approximately one third of the flocks (Fig. 2d, left). The probability that the same agent led as in the previous flock of the same trial was 1.0 when fewer than 50 steps separated the collapse of the previous flock from the formation of the next, and approximately 0.6 for 50–500 steps. When the gap exceeded 500 steps, this probability returned to the chance level of 1/3 (Fig. 2d, right).

**C. Cut links and phantom targets**

Figure 3 shows the changes around the cut for the 1,059 cut links of the leaders aligned at the time of the cut. Before the cut, the learning rate of these links had a median of approximately 0.4, the variance ratio $R/P$ was approximately 4, and the nullification strength had a median of about 0.4 (Fig. 3a, b).

Almost every analyzed cut (99%) arose from two consecutive resets, as did 92% of all 7,792 cut links. A reset one step before the cut was selected when the prediction error was small (median $\delta^2/P$ of 1.4) and drew a small likelihood variance (median $3.4 \times 10^{-4}$); the learning rate became $K \approx 1$, and the belief variance $P$ shrank to approximately $R$. Because $P$ had become small, the prediction error at the next step was large relative to the predictive variance, and a reset drew an $R$ of ordinary size (median 1.4). Then $R/P \approx R(0)/R(-1) \approx 3.7 \times 10^3$ (median), and the learning rate dropped to $K \approx 0$. This process produces a link cut.

As the cut continues, $R/P$ remains almost constant: from Eq. (7), $\tilde{R}^{t+1}/P^{t+1} = 1 + R^t/P^t$, so the ratio increases by only 1 per step regardless of β. Only a reset can reduce this ratio.

After the cut, the leader's turning angle rapidly decreased from a median of about 60° to about 2° after 20 steps and approximately 1° after 40 steps (Fig. 3c). The alignment $A$ increased from approximately $-0.1$ to about 0.9 after 10 steps and to approximately 0.97 after 40 steps (Fig. 3d). Meanwhile, the leader's belief error regarding the partner of the cut link increased to a median of approximately 55 after 70 steps and then leveled off as the phantom targets of shorter cuts began to return (Fig. 3e).

Figure 4 shows an example of a cut. After the cut, the leader's (agent 0's) belief about agent 1 is extrapolated with hardly any correction from observations and moves ahead, away from the actual position of agent 1 (Fig. 4b, dashed line). The leader moves straight toward this phantom target, and the two followers trail behind it (Fig. 4a). The flock was detected 76 steps after the link cut; 93 steps after the cut, the phantom target lay 69 units ahead of the leader. Note that cuts mostly began at short distances (across all 7,731 completed cuts, the median distance between the two agents at the onset of the cut was 1.4, and 90% were 6.6 or less), and the distance increased after the cut (for cuts lasting 100 steps or more, the median distance at onset was 1.5 and the median maximum distance during the cut was 8.8). Thus, cuts do not occur because the agents are far apart; instead, the distance increases after the cut.

**D. Mutual reinforcement of following and the leader's straight motion**

The role of following in maintaining a flock was examined using intervention experiments in which the system state was duplicated and only the behavior of either the followers or the leader was changed (Appendix C). When the followers stopped, the median time until the leader turned back decreased from 60 to 21 steps; conversely, when the leader kept moving straight, the other two agents continued to follow. With nearby followers, the leader's backward pull remains small, and the leader's straight motion and the followers' motion reinforce each other. However, this mutual reinforcement arises from the movement rule of the flocking model and is independent of nullification.

The link-cut duration (used here as a proxy for phantom-target lifetime) is strongly associated with $R/P$ at the cut (Fig. 5a; log–log slope 0.48, $r = 0.70$). This can be explained as follows. In a cut link, the estimated position is extrapolated at a constant velocity (Eq. (2)) and receives only small corrections from the learning rate $K$ (Eq. (3)). When the partner moves at nearly constant velocity, the belief error $d$ obeys

$$d^t - 2d^{t-1} + d^{t-2} \approx -Kd^{t-1} \qquad (13)$$

and behaves approximately as an undamped spring (harmonic oscillator) with angular frequency $\sqrt{K}$ (see Appendix B for the derivation). Substituting the learning rate at the cut, $K \approx P/R$, gives a half period of $\pi\sqrt{R/P}$, which most measured lifetimes follow closely (Fig. 5a, solid line); for the leader cuts that recovered within 2,000 steps, the median lifetime was 0.97 times this half-period, although some cuts lasted much longer. Moreover, the time course of the belief error normalized by its maximum closely follows $\sin(\pi x)$ when time is normalized by the half-period (Appendix B). Therefore, the phantom target swings forward and then returns.

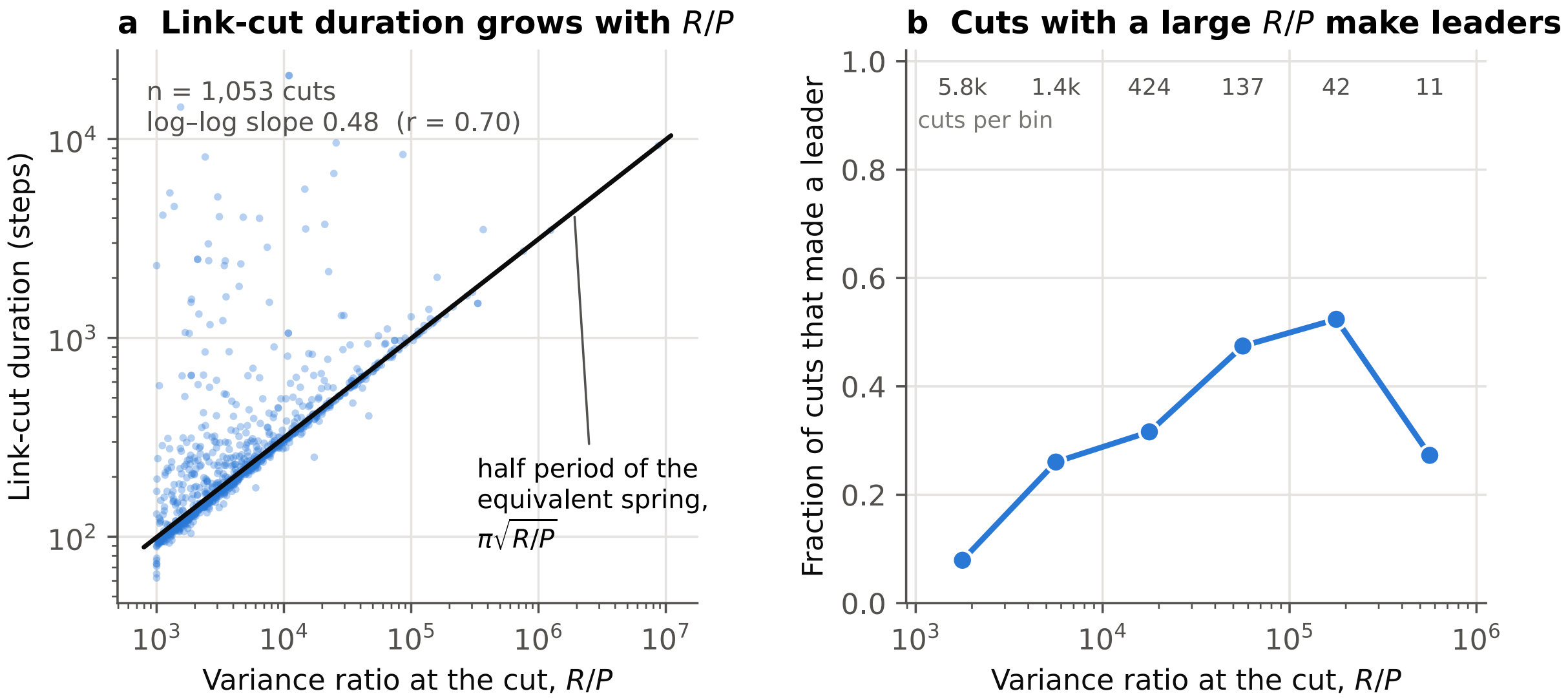

**FIG. 5.** Link-cut duration and phantom-target dynamics. (a) Link-cut duration (steps from the cut to recovery) versus R/P at the cut (n = 1,053 cuts that recovered before the end of the run; log–log slope 0.48, r = 0.70); line, half period of the equivalent spring, $\pi\sqrt{(R/P)}$ (Appendix B). (b) Fraction of cuts that produced a flock leader, by R/P at the cut (all 7,792 cut links whose onset fell within the analysis period, including 61 that had not recovered by its end; half-decade bins; number of cuts per bin at the top).

## E. Link-cut duration and its origin

From the cut mechanism described above, $R/P$ at the cut is nearly equal to the ratio of the two likelihood variances drawn in the two consecutive resets. The complementary cumulative distribution of $R/P$ over all 7,792 cut links whose onset was observed (the 7,731 completed cuts and 61 cuts that had not recovered by the end of the analysis period) had a tail slightly steeper than that of the ratio of two independent uniform random numbers, $T/x$ (exponent 1; estimated exponent 1.17; Appendix E). The 1,059 cuts that produced leaders were biased toward a large $R/P$ (median $3.7 \times 10^3$, versus $1.7 \times 10^3$ for all cuts).

The fraction of cuts that produced a leader increased with $R/P$, from 0.08 at $R/P \approx 2 \times 10^3$ to about 0.5 at $R/P \approx 10^5$ (Fig. 5b). Therefore, cuts that produce longer-lived phantom targets are more likely to produce leaders.

## F. Flock collapse

A flock whose leader holds a cut link could collapse in two ways: type A, in which the cut link recovers in the same step as the collapse, and type B, in which the link recovers at least one step after the collapse. Of the 1,043 such collapses, 1,040 were of type B and only 3 of type A.

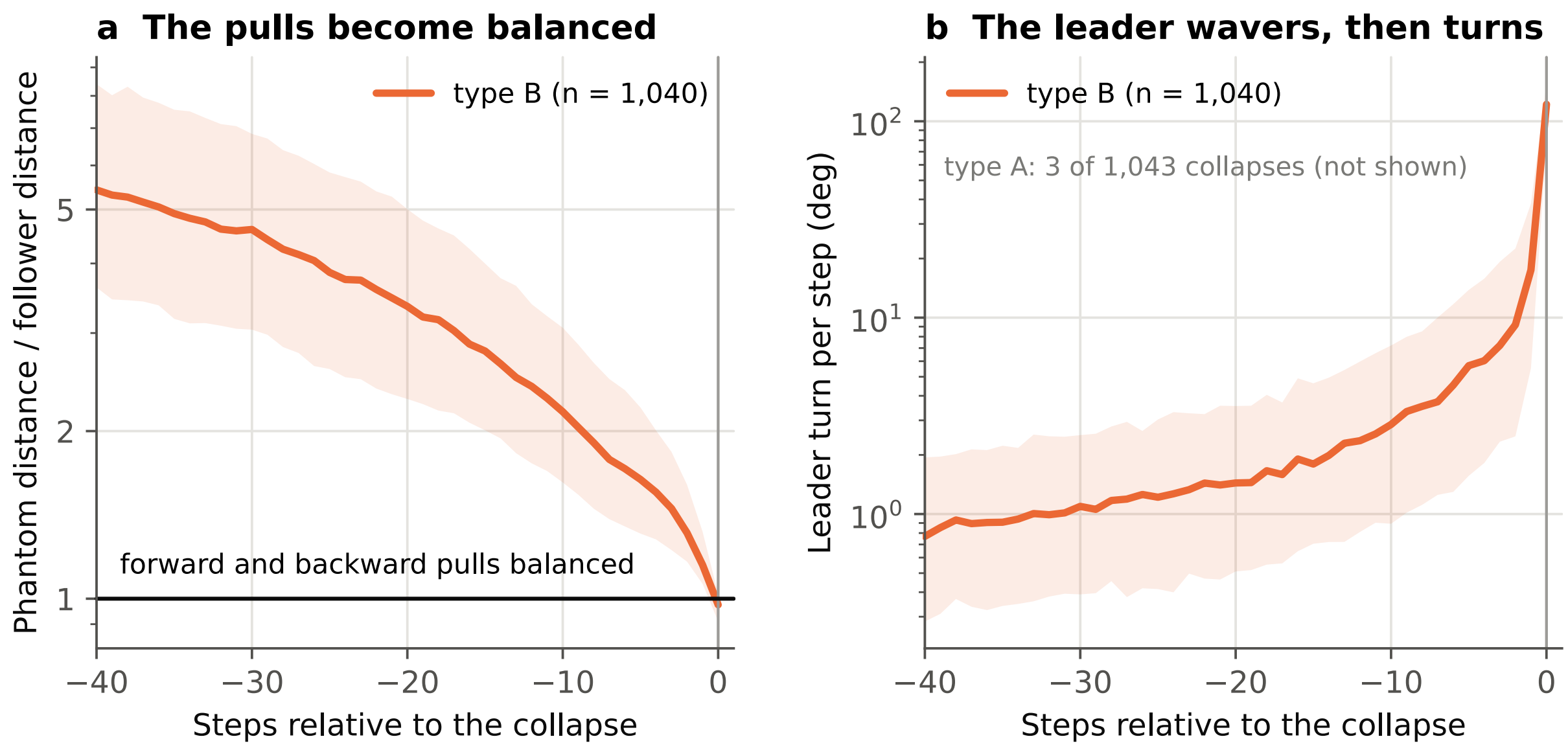

**FIG. 6.** Flock collapse (n = 1,040 collapses of type B, in which the cut link recovers at least one step after the collapse; type A, in which it recovers at the collapse step, occurred in only 3 of 1,043 collapses and is not shown). (a) Ratio of the leader's distance to the phantom target to its distance to the predicted position of the uncut follower before the collapse (median and interquartile range); the ratio falls to 1, where the forward and backward pulls balance. (b) Leader's turn per step (median and interquartile range): the leader wavers increasingly and then turns.

In type B, the cut was still continuing at the collapse (median $R/P$ of $4.6 \times 10^3$), and recovery occurred a median of 17 steps after the collapse (interquartile range 9–39). Before the collapse, the ratio of the distance to the phantom target to the distance to the predicted position of the follower decreased from a median of about 5 at 40 steps before the collapse and reached 1 at the time of collapse (Fig. 6a). The distance to the predicted position of the follower remained nearly constant (approximately 8), so the decrease in the ratio was due to the shortening distance to the phantom target, that is, the return of the phantom target (Appendix B). In Eq. (4), the leader's heading is determined by the sum of the forward vector toward the phantom target and the backward vector toward the follower whose link is not cut. In this nearly collinear configuration, the two vectors point approximately in opposite directions; thus, when their magnitudes become comparable, they nearly cancel. Type A collapses are rare because a reset that ends a cut is selected only when the prediction error has become small relative to the expanded variances [Eq. (12)], that is, when the phantom target has returned close to the partner (Sec. III G). By then, the phantom target has already passed the point at which the forward and backward pulls balance, so the flock collapses before the link recovers.

Correspondingly, the leader's turns fluctuated increasingly toward the collapse, from about 1° per step 40 steps before the collapse to about 17° one step before (Fig. 6b).

### G. Persistence and recovery of cuts

Table III summarizes the flocks and link cuts for each model. In zero-β BIB, which lacks nullification, the belief variance only decreases, and the learning rate decayed to about $7 \times 10^{-6}$ (median); redrawing the likelihood variance at a reset does not lower $R/P$ sufficiently, and almost no cuts recovered within the analysis period (51 recoveries; every link was still cut at the end of every trial). In 55 of the 100 trials, the agents remained several hundred units apart; in 40 trials, they stayed within a few units of one another but reversed their headings almost every step (Fig. 1e). The 932 flocks that formed were short (median 39 steps; 0.43% of the analysis period) and lacked an identifiable leader (the three criteria agreed in 36% of eligible flocks). In fixed-R BIB, which does not update the likelihood variance, the variance ratio shrinks whenever nullification occurs. Therefore cuts, although they occurred, recovered within a median of two steps (and at most five steps), and no flocks formed. In contrast, 1,603 flocks formed in the BIB, and cuts lasted a median of 96 steps, with 48% lasting at least 100 steps.

Cuts also occurred in fixed-β BIB ($\beta_0$ = 0.5) and random-R BIB, but they recovered within at most 31 and four steps, respectively, and no flocks formed. In the former, belief variance expanded rapidly, whereas in the latter, the variance ratio was not preserved. In FB, which has no resets, no link cuts occurred, and its flocks arose from translation (Fig. 1d).

For the 1,021 leader cuts in BIB that recovered within 2,000 steps of the cut, the belief variance expanded by a median factor of $3.4 \times 10^3$ by the time of recovery. While the phantom target swung forward, nullification also expanded the likelihood variance, keeping $\delta^2/R$ at about 1. Then, as the phantom target returned, the prediction error shrank but the expanded variances did not, so $\delta^2/R$ fell to a median of 0.0023 at recovery. Every recovery occurred through a reset (in 74% of cuts, the only reset during the cut); the variance ratio $R/P$ fell to a median of 1.1, and the learning rate rose to about 0.5. These cuts lasted a median of 0.97 times the half-period $\pi\sqrt{R/P}$; that is, recovery followed the return of the phantom target.

TABLE III. Flocks and cuts in each model (100 trials)

| Model | Flocks | Cuts | Cut length (median, max) | Fraction of time with a cut link |
|---|---|---|---|---|
| BIB | 1,603 | 7,731 | 96, 28,041 | 0.24 |
| fixed-β BIB | 0 | 7,458 | 14, 31 | 0.009 |
| random-R BIB | 0 | 9,614 | 2, 4 | 0.0015 |
| FB | 21,121 | 0 | — | 0 |
| zero-β BIB | 932 | — | — | 1.00 |
| fixed-R BIB | 0 | 5,076 | 2, 5 | 0.0008 |

Flock values are totals over 100 trials. Cuts are completed cuts, whose onset and recovery both fell within the analysis period (Sec. II G 2); cut lengths are measured in steps. In zero-β BIB, almost no cuts recovered within the analysis period (51 recoveries; every link was still cut at the end of every trial), so the number and length of cuts were not tallied using the criteria applied to the other models.

## IV. DISCUSSION

### A. Emergence of leaders through inference

We have shown that flocks with leaders form spontaneously, even though no alignment rule or leader-designating rule is specified, and all agents follow identical inference and movement rules. The three criteria (front agent, agent with a stable heading, and formation initiator) agreed in 79% of flocks (Fig. 2a), and the leader was not fixed to a particular agent (Fig. 2d). Thus, agents endogenously differentiate into distinct roles because of differences in their inference histories and internal states.

Collective motion without explicit alignment rules has already been reported in models based on mutual anticipation [24], active inference [28], and allocentric bearings [29]. In contrast to these and the other flocking models cited in the Introduction, this study has demonstrated a mechanism by which the persistence of predictions detached from observations produces the differentiation between leader and follower roles, and by which their resolution produces the collapse of flocks. Here, the leader is determined neither by information given in advance [33] nor by differences between individuals, such as in flight speed [31] or energetic state [34]; instead, it plays a dynamic role arising from inference and interindividual interaction.

### B. Role of nullification: reconciling the persistence and recovery of link cuts

The direct trigger of a link cut was not an increase in nullification strength, but a sharp rise in the variance ratio $R/P$ across two consecutive resets (Fig. 3). In contrast, the persistence and eventual recovery of a cut involve nullification, which acts with the same strength on both the belief and the likelihood. Because nullification expands the belief and likelihood variances by the same factor relative to the standard Bayesian update, it can increase both variances without changing their ratio. Because the nullification strength was sufficiently small during a link cut (90th percentile 0.011; Appendix B), the learning rate remains extremely low even when nullification occurs. While the phantom target swings forward, nullification widens the variances in step with the prediction error. When the phantom target returns, the widened variances become excessive relative to the prediction error, and a reset is selected. Therefore, the phantom target neither vanishes immediately nor persists indefinitely.

Comparison with the control models showed that without nullification, the learning rate decayed toward zero and cuts almost never recovered (zero-β BIB), and that without the likelihood side of nullification cuts resolved within a short time (fixed-R BIB). Cuts also did not persist long when the nullification strength was fixed at 0.5 or when the likelihood variance was replaced by random numbers (fixed-β BIB, random-R BIB; Table III).

In short, resets create link cuts, whereas nullification allows the cuts to persist while enabling their eventual recovery. In other words, nullification preserves the relative precision of belief and observation while lowering their absolute precision. The former sustains the phantom target, and the latter prepares its end as the phantom target returns. The seemingly opposing persistence of cuts and their recovery arises from two aspects of this single operation. These properties arise from the inference dynamics rather than from the flocking movement rule.

**C. Mechanisms specific to the flocking model**

In addition to the properties of inference described in the previous section, the prediction and movement rules of our model contribute to the emergence of leaders and the collapse of flocks. A link cut corresponds to the state described in the Introduction, in which the prediction is not substantially revised even when the observations and predictions diverge greatly. Because the predicted position continues to be extrapolated at constant velocity during a cut, it becomes a phantom target detached from reality, and the heading of the agent moving toward it stabilizes (Fig. 3c, d). The error between the predicted and actual positions approximately follows a spring-like oscillation, because the nullification strength is small during a cut and the update rule for the variance ratio is the same as in standard Bayes. As a result, the lifetime of the phantom target is roughly proportional to the square root of the variance ratio at the cut (Fig. 5a; Appendix B). Because this variance ratio is broadly distributed, roughly as the ratio of two uniform random numbers, long-lived phantom targets occasionally arise, and these are more likely to produce leaders (Fig. 5b; Appendix E).

The leader's heading is determined by the sum of the attractions toward the phantom target ahead and the followers behind it (Eq. (4)). When the followers remain close, the backward attraction is small, so the leader's straight motion is maintained, and this straight motion in turn encourages continued following (Appendix C). The same summation also underlies collapse. As the phantom target returns, it balances the forward and backward attractions and makes the heading unstable (Fig. 6); the cut link recovers only afterwards, once the prediction error has become small enough for a reset to be selected. Although nullification is not directly involved in the collapse itself, the collapse requires neither external commands nor explicit collapse rules, and this combination of the properties of inference and the movement rule supports the repeated formation and collapse of flocks with leaders.

**D. Theoretical significance of nullification**

Our results extend the role of nullification from individual inference to the formation of collective motion. Within the predictive processing framework, the importance of precision in weighting prediction errors has been emphasized [35, 36]. In our model, the learning rate depends on the relative precision of beliefs and observations. From this perspective, nullification lowers both precisions while preserving their ratio, and decouples the weakening of confidence in the existing belief from the strong incorporation of new observations, as discussed in the Introduction.

Indeed, during the leaders' link cuts, belief variance expanded greatly while the learning rate remained extremely small: confidence in the existing belief weakened, yet strong, immediate incorporation of new observations was avoided. This state can be regarded as one in which the attribution of the prediction error to the existing belief or to the new observation is not settled and its interpretation is temporarily suspended. This suspension is resolved by a reset when the phantom target returns and the widened variances become excessive relative to the prediction error. The likelihood variance redrawn at this point is comparable to the expanded belief variance (the variance ratio falls to a median of 1.1), so the learning rate rises from about $10^{-4}$ to about 0.5. The resulting update revises the belief, removing the phantom target as observations are incorporated anew.

Importantly, this inferential state is not confined to the individual but affects collective motion through interactions with others. In other words, the accuracy of estimation at the individual level and the emergence of order at the collective level do not necessarily coincide. As the control models showed, the formation of collective order requires not only a discrepancy between prediction and reality but also that this discrepancy persists for a certain period and is subsequently resolved.

The decoupling described above distinguishes nullification from the adaptive-learning schemes cited in the Introduction. Forgetting methods weaken the influence of past information [2], whereas adaptive Bayesian learning models adjust learning rates according to inferred environmental change, volatility, or relative uncertainty [3–5]. Changepoint detection [6], context-dependent learning in the presence of changepoints or oddballs [7], and the joint estimation of volatility and stochasticity [8] instead infer whether errors reflect a change in the state or noise, and adjust learning accordingly. Nullification follows neither route: it weakens confidence in the existing belief while keeping the learning rate low, and defers the attribution of the error until a reset is selected. The link cuts and phantom targets observed here arise from this deferred state. When only the belief variance was expanded, so that weakening confidence in the existing belief also raised the learning rate, cuts either did not occur (FB) or resolved within at most five steps (fixed-R BIB), and the characteristic leader–follower flocks of BIB did not emerge (Fig. 1; Table III).

**E. Limitations and future directions**

This study used a minimal model with three agents in which all individuals always observe one another and move without inertia. In addition, a one-dimensional BIB update rule was applied to two-dimensional position estimation, and the results are limited to the condition $R_0 = 1$. Drawing the likelihood variance at a reset from a uniform distribution is a modeling assumption. As shown in Appendix E, the frequency with which sufficiently small likelihood variances are drawn determines the frequency of flocking and how clearly a leader can be identified. In particular, when the likelihood variance drawn at a reset was bounded below by $2\times10^{-3}$ times the reference likelihood variance, only 79 flocks formed, and distributions with many values near zero kept the links cut almost permanently (Appendix E 2). In addition, the reset criterion in Eq. (12) uses the pre-update predictive distribution, which, unlike the updated predictive distribution, does not favor very small likelihood variances. When the criterion was instead evaluated with the updated predictive distribution and the nullification strength was set to zero at a reset, leader–follower flocks also formed (2,716 flocks; the three criteria agreed in 87% of eligible flocks), although cuts were shorter (Appendix E 3). The present results therefore concern a reset-based BIB model in which sufficiently small likelihood variances can be drawn. Furthermore, because the magnitude of attraction toward a predicted position is proportional to the predicted distance, distant phantom targets strongly influence the heading. The dependence on parameters such as the reference likelihood variance and the attraction rule remains to be examined.

Moreover, the observation noise in our model is Gaussian, and the model includes neither a separate process that generates outliers nor nonstationary observation noise. Thus, although nullification can be regarded as a temporary suspension of the interpretation of prediction errors, the present collective model does not directly demonstrate its adaptive advantage in situations where state changes and outliers are mixed. Future work should examine whether the same mechanism of leader formation holds in environments with such ambiguous observations, in larger groups, and under different interaction conditions, such as a limited field of view.

Finally, in stickleback pairs, individuals that are less responsive to their partner's movements have been reported to become leaders, and this role is reinforced by social feedback [37]. In our model, agents with reduced responsiveness (link cuts) become leaders, and their straight movements and the followers' responses reinforce each other. Here, this reduced responsiveness is a transient state produced by inference. However, if individual differences are introduced into the parameters that determine how readily link cuts occur (such as the distribution of the likelihood variance at a reset), leadership based on individual differences could also be treated within the same framework.

## V. CONCLUSIONS

We have shown that three agents that use BIB inference to predict one another's positions and move toward those predictions repeatedly form leader–follower flocks that subsequently collapse despite having no explicit alignment rules or designated leaders. When link cuts suppress correction by observations, predicted positions detached from reality (phantom targets) form; the straight motion of the agent heading toward such a target and the responses of the other agents produce group alignment and reinforce one another, maintaining the flock. Flocks collapsed when the forward and backward attractions balanced as the phantom target returned, and the cut link recovered shortly afterwards.

The link cuts that triggered flock formation arose from sharp increases in the variance ratio across consecutive resets. By widening the belief and likelihood variances by the same factor, nullification preserved relative precision, allowing the cuts to persist, while preparing their subsequent recovery at a later reset by lowering absolute precision. Nullification thus reconciles the seemingly opposing processes of persistence and recovery of cuts within a single operation, thereby contributing to both the maintenance and collapse of flocks. These results indicate that, in this model, predictions inaccurate at the individual level can give rise to collective order, and they provide a mechanistic account of how individual inference can lead to emergent leadership and the repeated formation and collapse of flocks.

These findings also suggest directions beyond the present model. In multi-agent tracking and swarm robotics, some adaptive estimators raise their gain in response to large innovations, which can make agents overreact to transient outliers. Nullification offers an alternative in which a temporary mismatch between prediction and observation is tolerated without abrupt revision while recovery remains possible; in the present model, the same mechanism generated transient, self-organized leadership without leaders designated in advance. Testing whether these properties hold in larger groups, under limited fields of view, in environments that mix state changes and outliers, and in physical robot swarms is an important direction for future work.

## ACKNOWLEDGMENTS

This work was supported by JSPS KAKENHI Grant No. JP21K12009. During the preparation of this manuscript, the authors used Claude Opus 5.5 (Anthropic) and ChatGPT (GPT-6; OpenAI) to assist in translating the original Japanese manuscript into English and improving the clarity and quality of the English text. The simulation code was written by the authors; AI tools assisted in extending it for the intervention experiments (Appendix C) and in developing the analysis and plotting code. All AI-assisted output was reviewed and verified by the authors, who take full responsibility for the final content of the manuscript. We would like to thank Editage (www.editage.jp) for English language editing.

## DATA AVAILABILITY

The code used to generate the data in this article will be deposited in a public repository with a persistent identifier upon publication. All simulation results reported here can be fully reproduced with this code.

## AUTHOR CONTRIBUTIONS

S. S. contributed to conceptualization, formal analysis, methodology, software, investigation, validation, funding acquisition, and writing – original draft. D. M. and Y. N. contributed to conceptualization and writing – review & editing. T. T. and M. H. contributed to writing – review & editing. U. C. and Y.-P. G. contributed to supervision and writing – review & editing.

## APPENDIX A

As in Sec. II D, the direction of the prediction-error vector is treated as a one-dimensional axis with the predicted position as the origin. Omitting the time index, we write the belief distribution, the likelihood and the predictive distribution as

$$q(s) = N(s; 0, P), \quad p(o \mid s) = N(o; s, R), \quad p(o) = N(o; 0, P + R) \quad (A1)$$

Because $\Lambda(s, o) = p(o \mid s)/p(o)$, the nullification kernel in Eq. (6) can be written as:

$$K_\beta(s, o) = q(s)^{1-\beta} p(o)^{-\beta} p(o \mid s) \qquad (A2)$$

### 1. Belief update

We fix $o$ at the observed value $\delta$ and normalize over $s$. Because $p(o)^{-\beta}$ does not depend on $s$, we obtain

$$q'(s) \propto q(s)^{1-\beta} p(\delta \mid s) \propto exp\left[-\frac{(1-\beta)s^2}{2P} - \frac{(\delta - s)^2}{2R}\right] \quad (A3)$$

Completing the square in $s$ shows that $q'(s)$ is a Gaussian distribution with mean $K\delta$ and variance $P'$, where

$$K = \frac{P}{P + (1-\beta)R}, \qquad P' = \frac{PR}{P + (1-\beta)R} \quad (A4)$$

These results agree with the updates of the learning rate and belief variance in Eq. (7). The shift of the estimate from the predicted position by $K\delta$ in the direction of the error corresponds to Eq. (3). Moreover, since $q(s)^{1-\beta}$ is proportional to a Gaussian distribution with variance $P/(1-\beta)$, this update is equivalent to inflating the belief variance by a factor of $1/(1-\beta)$ and then performing a standard Bayesian update. This has the same form as exponential forgetting [2].

### 2. Likelihood update

We fix $s$ and normalize over $o$. Since $q(s)^{1-\beta}$ does not depend on $o$, we obtain

$$r(o \mid s) \propto p(o)^{-\beta} p(o \mid s) \propto exp\left[\frac{\beta o^2}{2(P+R)} - \frac{(o - s)^2}{2R}\right] \quad (A5)$$

From the quadratic coefficient in $o$, $r(o \mid s)$ is a Gaussian distribution with variance

$$\tilde{R} = \left[\frac{1}{R} - \frac{\beta}{P + R}\right]^{-1} = \frac{R(P + R)}{P + (1-\beta)R} \qquad (A6)$$

For $0 \leq \beta \leq 1$, the right-hand side is positive, and it agrees with $\tilde{R}^{t+1}$ in Eq. (7). The mean of $r(o \mid s)$ is $s(P + R)/[P + (1-\beta)R]$, but in our model, as in [12], the center of the likelihood at the next time step is kept at $s$, and only the variance $\tilde{R}$ is carried over.

### 3. Preservation of the coupling structure and normalizability

Taking the logarithm of (A2) gives

$$log K_\beta(s,o) = -\frac{(1-\beta)s^2}{2P} + \frac{\beta o^2}{2(P+R)} - \frac{(o-s)^2}{2R} + const. \quad (A7)$$

Expanding this expression shows that the coefficient of the cross term in $s$ and $o$ is $1/R$, which does not depend on the nullification strength $\beta$. Hence, the nullification kernel changes the coefficients of the quadratic terms in the state and in the observation while preserving their cross term.

This property implies that the cross odds ratio for any two states $s_1, s_2$ and two observations $o_1, o_2$ does not depend on $\beta$:

$$\frac{\mathcal{K}_\beta(s_1,o_1)K_\beta(s_2,o_2)}{K_\beta(s_1,o_2)K_\beta(s_2,o_1)} = exp\left[\frac{(s_1-s_2)(o_1-o_2)}{R}\right]. \quad (A8)$$

In this sense, the nullification kernel changes the probabilistic weighting of the state and the observation while preserving their cross-coupling structure. However, because our model imposes the additional constraint of fixing the center of the likelihood in the likelihood update, this invariance does not carry over unchanged to the updated joint distribution.

For $\beta = 0$, (A2) coincides with the original joint distribution $p(s,o)$, and (A4) and (A6) reduce to the standard Gaussian Bayesian update ($\tilde{R} = R$).

Furthermore, for $0 \le \beta \le 1$, the necessary and sufficient condition for the nullification kernel $\mathcal{K}_\beta(s,o)$ to be integrable over $(s,o)$ is

$$(1-2\beta)P + (1-\beta)^2 R > 0 \quad (A9)$$

For $P, R > 0$, this condition is equivalent to

$$0 \le \beta < \frac{\sqrt{P+R}}{\sqrt{P+R}+\sqrt{P}} \quad (A10)$$

Hence, in general, the nullification kernel can be normalized as a joint probability distribution only within this range. In contrast, the conditional distributions (A3) and (A5) used in the actual updates are well defined as Gaussian distributions over the entire range $0 \le \beta \le 1$.

## APPENDIX B: LIFETIME OF THE PHANTOM TARGET: SPRING APPROXIMATION OF THE BELIEF ERROR

We derive Eq. (13) in the main text and the lifetime of the phantom target, $\pi\sqrt{R/P}$. For a cut link, the agent indices $ij$ are omitted and only the time index is kept.

### 1. Recurrence for the belief error

Let $\mathbf{x}^t$ be the true position of the partner, $\mathbf{o}^t = \mathbf{x}^t + \boldsymbol{\varepsilon}^t$ the observation and $\mathbf{d}^t = \mathbf{m}^t - \mathbf{x}^t$ the belief error. Because the learning rate $K^t$ is a scalar that multiplies the two-dimensional vector directly, the following equations hold for each component of the vector. From Eqs. (2) and (3),

$$\mathbf{m}^t = (1-K^t)\,\hat{\mathbf{m}}^t + K^t\mathbf{o}^t, \qquad \hat{\mathbf{m}}^t = 2\mathbf{m}^{t-1} - \mathbf{m}^{t-2} \quad (B1)$$

Substituting $\mathbf{m} = \mathbf{x} + \mathbf{d}$ and subtracting $\mathbf{x}^t$ gives

$$\mathbf{d}^t = (1-K^t)(2\mathbf{d}^{t-1} - \mathbf{d}^{t-2}) - (1-K^t)\,\mathbf{a}^t + K^t\boldsymbol{\varepsilon}^t \quad (B2)$$

where $\mathbf{a}^t = \mathbf{x}^t - 2\mathbf{x}^{t-1} + \mathbf{x}^{t-2}$ is the acceleration of the partner.

### 2. Approximations during a cut

Three approximations were made for each link cut. First, the learning rate is almost constant and can be regarded as $K^t \approx K \approx P/R$. Over all steps of the leaders' cuts, the nullification strength was zero or very small at most steps (90th percentile 0.011), and the median of $K \cdot R/P$ was 1.000. Moreover, since $R/P$ increases by only 1 per step under the nullification update, for cuts with $R/P \sim 10^3$–$10^4$ the changes in $R/P$ and $K$ over the lifetime (hundreds of steps) are small. Indeed, the change in $K$ at 300 steps relative to 10 steps after the cut was about 2% in the median. Second, the partner moves straight at a nearly constant speed, so we set $\mathbf{a}^t \approx \mathbf{0}$. Third, since $K \sim 10^{-4}$, the observation-noise term $K\boldsymbol{\varepsilon}^t$ ($\sigma = 1$) can be neglected. Then, (B2) becomes

$$\mathbf{d}^t = (1-K)(2\mathbf{d}^{t-1} - \mathbf{d}^{t-2}) \quad (B3)$$

Subtracting $2\mathbf{d}^{t-1} - \mathbf{d}^{t-2}$ from both sides gives:

$$\mathbf{d}^t - 2\mathbf{d}^{t-1} + \mathbf{d}^{t-2} = -K\,\mathbf{d}^{t-1} - K(\mathbf{d}^{t-1} - \mathbf{d}^{t-2}) \quad (B4)$$

The second term on the right-hand side represents the change in the error per step multiplied by $K$. As shown below, this change is on the order of $\sqrt{K}$ times the error, so the ratio on the second term to the first is of the order of $\sqrt{K}$. Neglecting the second term yields Eq. (13) in the main text.

### 3. Solution and lifetime

(B3) is a linear recurrence with constant coefficients, and the roots of its characteristic equation $\lambda^2 - 2(1-K)\lambda + (1-K) = 0$ are

$$\lambda = \sqrt{1-K}\left(\sqrt{1-K} \pm i\sqrt{K}\right) = \sqrt{1-K}\ e^{\pm i\theta},$$
$$sin\theta = \sqrt{K} \quad (B5)$$

The error is small at the time of the cut, but because the estimated position jumps to the observation at the reset just before the cut ($K \approx 1$), the extrapolation velocity deviates from the partner's velocity. Denoting this deviation by $\Delta\mathbf{v}$ and using the initial conditions $\mathbf{d}^0 = \mathbf{0}$ and $\mathbf{d}^1 = \Delta\mathbf{v}$, we obtain

$$\mathbf{d}^t = \Delta\mathbf{v}\,(1-K)^{(t-1)/2}\,\frac{\sin(\theta t)}{\sin\theta} \qquad (B6)$$

The error vector swings forward and then returns while keeping the direction of $\Delta\mathbf{v}$. The half-period until the error returns to 0 is:

$$T = \frac{\pi}{\theta} = \frac{\pi}{arcsin\sqrt{K}} \approx \frac{\pi}{\sqrt{K}} \approx \pi\sqrt{\frac{R}{P}} \qquad (B7)$$

which approximates the lifetime of the phantom target (Fig. 5a). The maximum error is of the order of $|\Delta\mathbf{v}|/\sqrt{K} \approx |\Delta\mathbf{v}|\sqrt{R/P}$ . Over a half period the amplitude decays by a factor of $(1-K)^{T/2} \approx$

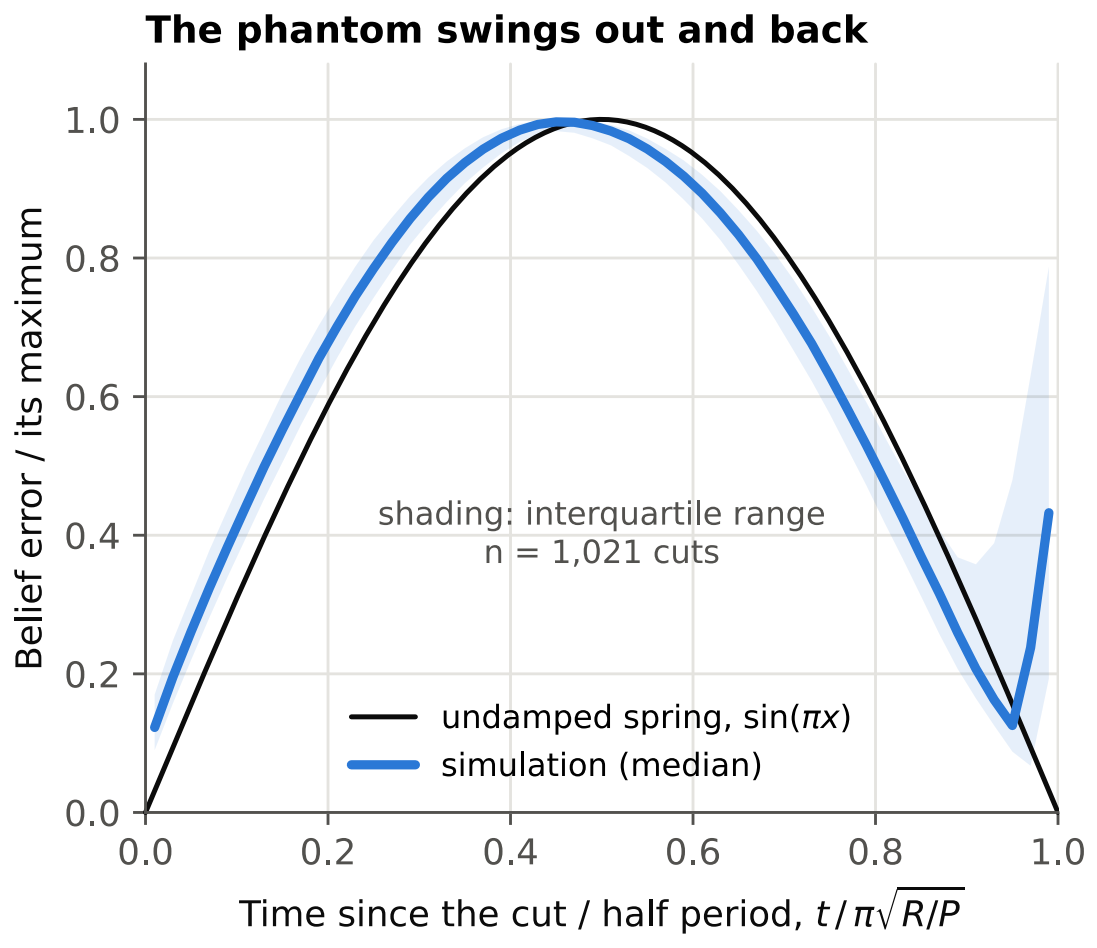

**FIG. 7.** Belief error, normalized by its maximum, plotted against time normalized by the half-period π√(R/P) of the equivalent spring (median and interquartile range; n = 1,021 cuts that recovered within 2,000 steps), compared with the undamped spring sin(πx).

$e^{-\pi\sqrt{K}/2}$, a reduction of only about 2% for $K \sim 10^{-4}$; this is why the spring was described as undamped in the main text. In addition, (B6) shows that when the error is normalized by its maximum and time by the half-period, the curves closely follow $\sin(\pi x)$ (Fig. 7).

## APPENDIX C: MUTUAL REINFORCEMENT OF FOLLOWING AND THE LEADER'S STRAIGHT MOTION

### 1. Methods

Using intervention experiments that duplicate the system, we examined whether the followers' motion and the leader's straight motion reinforce one another in maintaining a flock. We selected cases in which a cut link of an agent had lasted 150 steps and the alignment was A≥0.9 at that time (1,159 cases in 100 trials), and duplicated the entire state of the three agents at that time. In this appendix, the agent holding this cut link is called the leader (the three criteria were not applied). Because the duplicated system receives the same observation noise as the original, an unmodified duplicate reproduces the original exactly. In the duplicated system, exactly one of the following was changed. In "followers stopped", the two agents other than the leader neither infer nor move and stay in place. In "followers straight", the two agents other than the leader do not infer and keep moving straight at speed 1 in their headings at the time of duplication. In "leader straight", the leader does not infer and keeps moving straight at speed 1 in its heading at the time of duplication. The first time at which the leader's heading deviated by more than 90° from its heading at the time of duplication was taken as the time at which the leader turned back.

### 2. Results

When the followers were stopped, the mean distance between the leader and the other two agents increased to a median of 34 after 100 steps (3.4 without changes), and the leader turned back in less than half the time of the original system (median ratio for the same case 0.42; earlier in 92% of cases) (Table IV). When the followers kept moving straight without responding, they remained within about 15 units of the leader after 100 steps, and the time until the leader turned back hardly changed (ratio 1.00). Conversely, when the leader was kept moving straight, the other two agents continued to follow, and the fraction of the 5,000 steps in which A ≥ 0.9 was maintained had a median of 99.4% (note to Table IV). Hence, the leader's straight motion and the followers' motion support each other and contribute to maintaining the flock. The lifetime of a flock, however, also involves the persistence and return of the leader's phantom target and the recovery of the link by a reset.

TABLE IV. Results of the intervention experiments

| Condition | Steps until the leader turns back (median, IQR) | Ratio to unmodified | Mean distance after 100 steps (median) |
|---|---|---|---|
| Unmodified | 60 (20–156) | — | 3.4 |
| Followers straight | 65 (22–163) | 1.00 | 15 |
| Followers stopped | 21 (6–49) | 0.42 | 34 |
| Leader straight | — (forced straight) | — | 9.9 |

In “leader straight”, the leader does not turn back, so instead we computed for each case the fraction of the 5,000 steps with A ≥ 0.9. Its median was 99.4% (interquartile range 98.7–99.7%). Turn-back statistics exclude the cases in which the leader did not turn back within 5,000 steps (unmodified, 3; followers straight, 46; followers stopped, 1).

## APPENDIX D: FORMULATION OF THE CONTROL MODELS

In this section, the formulations of the five control models listed in Table I of the main text are presented. In all models, the rules for observation, prediction and movement (Eqs. (1)–(4)) are the same as in BIB.

### 1. Fixed-β BIB

In fixed-β BIB, the candidate nullification strength is not determined automatically but is set to a fixed value $\tilde{\beta}^t = \beta_0 = 0.5$. However, the reset criterion is applied, as in BIB. This control model examines the role of automatically determining the nullification strength based on the prediction error.

### 2. Random-R BIB

In random-R BIB, updating the candidate likelihood variance by Eq. (7) is replaced by a random setting. Specifically, the candidate likelihood variance for the next time step is $\tilde{R}^{t+1} = R' = 2R_0 u'$, $u' \sim U(0,1)$, where $u'$ is generated at each time step and for each link independently of the random number $u^t$ used for the reset criterion.

Because the next-step candidate likelihood variance in this model differs from that in Eq. (7), the predictive surprise used for the nullification strength is also recomputed for this candidate. With the updated predictive variance $P^{t+1} + R'$, the predictive surprise is

$$-\log\tilde{p}^{t+1}(\delta^t; R, \beta) = \frac{1}{2}\log\frac{2\pi N}{P + (1-\beta)R} + \frac{(1-\beta)^2 R^2 \delta^2}{2\{P + (1-\beta)R\}N} \quad (D1)$$

where $N = P(R + R') + (1-\beta)RR'$. Replacing $R'$ with the candidate likelihood variance of Eq. (7) gives $N = R(2P + R)$, and Eq. (D1) then coincides with Eq. (8).

The reset criterion is Eq. (12), in which the carried-over candidate $\tilde{R}^t$ is the $R'$ drawn at the previous time step. The candidate nullification strength is determined by minimizing Eq. (D1) under the $R = R^t$ used at time $t$.

Its closed-form solution is

$$\tilde{\beta}^{t+1} = \begin{cases} 0, & G \le 0, \\ \max\{0,\, 1 - b^*\}, & G > 0, \end{cases} \quad (D2)$$

$$b^* = \frac{P\left\{B + \sqrt{B^2 + 4R\,(R + R')\,G}\right\}}{2RG} \quad (D3)$$

where $G = \delta^2(R + 2R') - RR'$ and $B = R(R + 2R') - 2\delta^2(R + R')$; on the right-hand side, we write $P = P^t$, $R = R^t$ and $\delta = \delta^t$. The learning rate and belief variance are updated using Eq. (7).

Thus, in random-R BIB, only the candidate likelihood variance is replaced by random numbers, and the nullification strength is based on the predictive distribution that reflects this candidate.

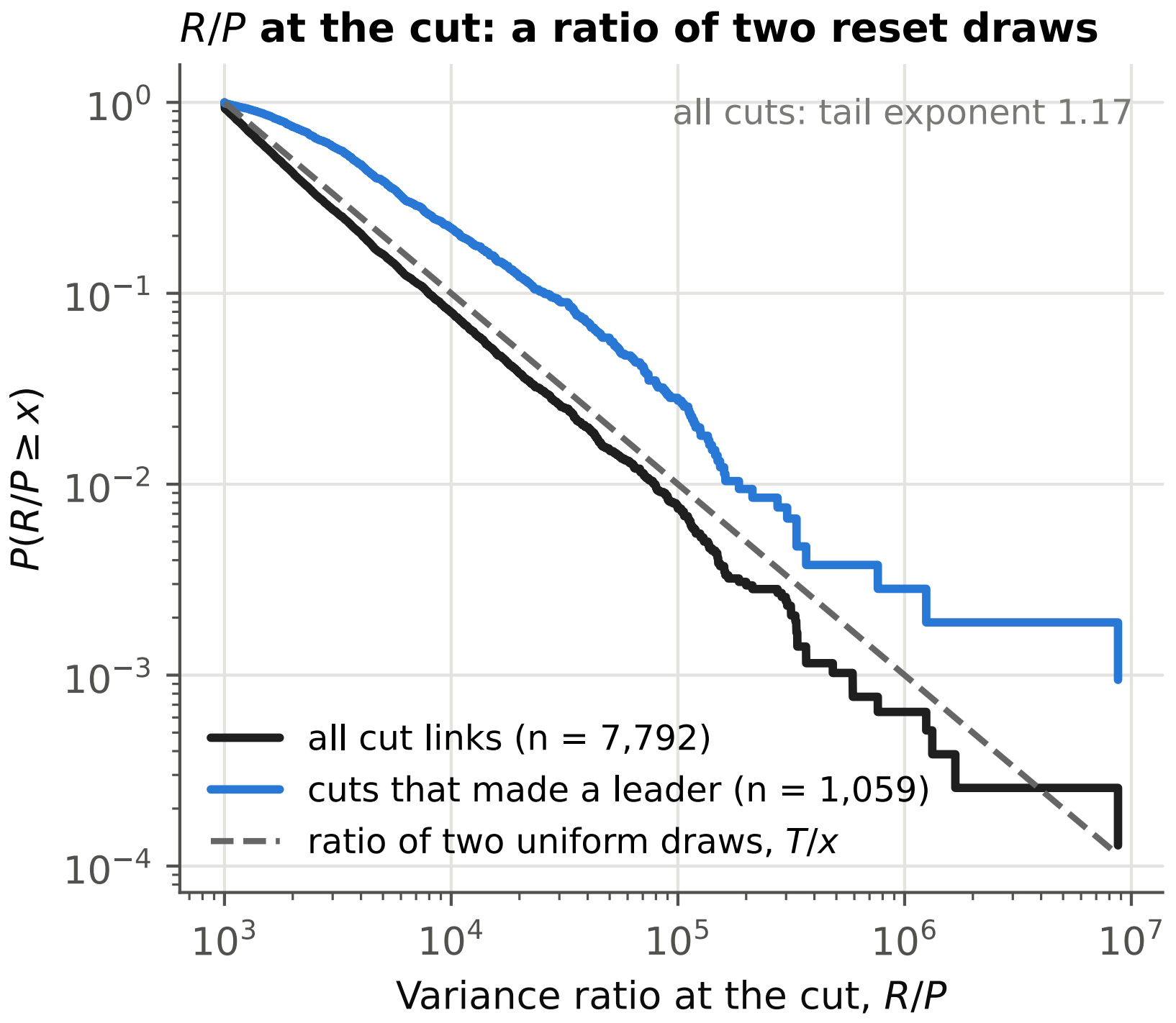

**FIG. 8.** Complementary cumulative distribution (CCDF) of R/P at the onset of every cut link whose onset fell within the analysis period (n = 7,792, including 61 that had not recovered by its end) and of the cuts that made a flock leader (n = 1,059). Dashed, T/x: the tail of the ratio of two independent U(0,1) draws above the cut threshold T = 1,000 (fitted tail exponent of all cuts, 1.17).

**3. FB**

FB (forgetting Bayes) is obtained from fixed-β BIB by additionally fixing the likelihood variance at $R_0$ and performing no resets. The nullification strength at each time step is fixed at $\beta^t = \beta_0$. Then $\beta_0$ acts as a forgetting rate that inflates the belief variance, giving $P^{t+1} = \frac{P^t R_0}{P^t + (1-\beta_0) R_0}$. The belief variance converges to $P^* = \beta_0 R_0$, at which point the learning rate is $K^* = \beta_0$. In this study, the forgetting rate of FB was set to $\beta_0 = 0.05$.

**4. Zero-β BIB**

The zero-β BIB model is a fixed-β BIB with $\beta_0 = 0$. In this case, the likelihood-variance update in Eq. (7) becomes $\tilde{R}^{t+1} = R^t$, and the belief update coincides with a standard Bayesian update. This control model retains the randomization of the likelihood variance at a reset and the reset criterion, and removes only nullification.

**5. Fixed-R BIB**

In fixed-R BIB, the nullification strength is determined automatically, but the likelihood variance is not updated, so $\tilde{R}^{t+1} = R^t$. The likelihood variance selected at a reset is kept until the next reset. Thus, the model keeps the update of the belief variance by nullification and removes only the likelihood side of the nullification (the update of the likelihood variance). The candidate nullification strength is obtained by setting $R' = R$ in Eqs. (D1)–(D3).

## APPENDIX E: ROBUSTNESS TO THE RESET DISTRIBUTION AND CRITERION

**1. Distribution of the variance ratio at the cut**

The variance ratio $R/P$ at a cut is nearly equal to the ratio $R(0)/R(-1)$ of the two likelihood variances drawn in the two consecutive resets (Fig. 3). If the two accepted values were independent draws of $2R_0 u$ ($u \sim U(0,1)$), this ratio would follow the ratio $X$ of two independent uniform random numbers. Because $X$ follows $\Pr(X > x) = 1/(2x)$ for $x \geq 1$, the complementary cumulative distribution above the cut threshold $T$ would be $T/x$. The complementary cumulative distribution of $R/P$ over all 7,792 cut links whose onset was observed had a tail slightly steeper than this one (estimated exponent 1.17; Fig. 8), presumably because each draw is accepted by the reset criterion depending on the prediction error. The 1,059 cuts that produced leaders were biased toward larger $R/P$ (median $3.7 \times 10^3$, versus $1.7 \times 10^3$ for all cuts).

**2. Distribution of the likelihood variance at reset**

To examine the role of small likelihood variances that trigger cuts, we ran simulations in which only the distribution of the likelihood variance at a reset was changed, using the same settings as in the main text (100 trials each; Table V). With a lower bound, $2R_0U(\varepsilon, 1)$, the maximum of $R/P$ at the onset of a cut was limited to about $1/\varepsilon$ ($9.0 \times 10^4$, $1.1 \times 10^4$ and $1.9 \times 10^3$ for $\varepsilon = 10^{-5}$, $10^{-4}$ and $10^{-3}$, respectively); flocks and leaders decreased with $\varepsilon$, and only 79 flocks formed for $\varepsilon = 10^{-3}$. Removing the values below $2 \times 10^{-5}R_0$, which account for only $10^{-5}$ of the values drawn at resets, hardly changed the number of flocks but reduced the fraction of time in flocks by about 20%. A distribution whose density approaches 0 near 0 ($u^{1/2}$) produced almost no flocks. Conversely, with distributions containing many values near 0 ($u^2$ and log-uniform), the links remained cut almost permanently (all six links were still cut at the end of every trial), few flocks formed, and their initiators could rarely be determined. In other words, for flocks with leaders to form, sufficiently small likelihood variances must be drawn at resets, but not so often that cuts become permanent; the lower bound and frequency of small values strongly affect how often flocks form and how clearly a leader can be identified. With the upper bound halved ($R_0u$), fewer flocks formed (1,104), whereas the agreement among the three criteria was unchanged (80%). In fixed-R BIB, which uses the same resets, cuts vanished within a few steps, even when they occurred at an $R/P$ comparable to that in BIB (Table III). Small likelihood variances trigger cuts, but the BIB update rule sustains phantom targets and produces leaders.

**3. Reset criterion**

The reset criterion in Eq. (12) evaluates the current prediction error with the pre-update predictive distribution $N(\delta^t; 0, P^t + R)$, whose density remains bounded as $R \to 0$. An alternative is to compare the predictive surprises of Eq. (8) after the update: the reset update $(2R_0u^t, 0)$ is then compared with the nullification update $(\tilde{R}^t, \tilde{\beta}^t)$, and the nullification strength is set to zero at a reset. Because the updated predictive density diverges as $R \to 0$, this post-update criterion favors very small likelihood variances. Flocks with leaders also formed with this criterion (Table V); flocks were more frequent (2,716) and the three criteria agreed more often (87%), but cuts were shorter (median 65 steps, versus 96 steps in the baseline). This is presumably because, with the pre-update criterion, a reset that ends a cut is selected only when the prediction error has become small, that is, when the phantom target has returned (Sec. III G). Setting the nullification strength to zero at a reset while keeping the pre-update criterion also produced flocks with leaders (2,042 flocks; agreement 76%). With the post-update criterion and $R_0u$ at a reset, which is the model of an earlier version of this study, cuts were shorter still (median 17 steps).

TABLE V. Simulations with different reset distributions and criteria (100 trials each)

| Criterion | $\beta$ at a reset | Likelihood variance at reset | Flocks | Flocks with a leader | Agreement of three criteria | Time in flocks | Cut length |
|---|---|---|---|---|---|---|---|
| Pre-update | Candidate | $2R_0u$ | 1,603 | 1,116 | 79% | 2.5% | 96 |
| Pre-update | Candidate | $2R_0U(10^{-5}, 1)$ | 1,601 | 1,130 | 79% | 2.0% | 92 |
| Pre-update | Candidate | $2R_0U(10^{-4}, 1)$ | 1,160 | 820 | 77% | 0.91% | 89 |
| Pre-update | Candidate | $2R_0U(10^{-3}, 1)$ | 79 | 54 | 73% | 0.033% | 35.5 |
| Pre-update | Candidate | $2R_0u^{1/2}$ | 2 | 2 | — | 0.0008% | 4 |
| Pre-update | Candidate | $2R_0u^2$ | 117 | 0 | —* | 3.4% | 174 |
| Pre-update | Candidate | Log-uniform | 59 | 0 | —* | 0.26% | 6 |
| Pre-update | Candidate | $R_0u$ | 1,104 | 772 | 80% | 2.2% | 106 |
| Pre-update | 0 | $2R_0u$ | 2,042 | 1,345 | 76% | 3.1% | 91 |
| Post-update | 0 | $2R_0u$ | 2,716 | 1,912 | 87% | 7.0% | 65 |
| Post-update | 0 | $R_0u$ | 1,615 | 1,187 | 90% | 5.3% | 17 |

Flocks are totals over 100 trials. Time in flocks is the fraction of the analysis period occupied by flocks, and cut length is the median number of steps. In the criterion column, "pre-update" denotes Eq. (12) and "post-update" the criterion using the updated predictive surprise of Eq. (8). In the $\beta$ column, "candidate" means that the candidate nullification strength is used at a reset (as in the main text), and "0" that it is set to zero. $u$ denotes a random number drawn from $U(0,1)$; $U(a, 1)$ denotes the uniform distribution on the interval $(a, 1)$; and log-uniform denotes the log-uniform distribution on $[2 \times 10^{-6}R_0, 2R_0]$. The first row is the baseline; the last row is the model of an earlier version of this study. The agreement of the three criteria was computed for flocks whose formation initiator was determined, whose low-turning interval began within the search window, and that collapsed before the end of the record (the same flocks as in Fig. 2a–c); a dash indicates that fewer than 10 flocks could be evaluated. In the baseline, the three criteria agreed for 1,116 of 1,411 flocks. *For $2R_0u^2$ and log-uniform, the initiator could not be determined for 115 of 117 and 58 of 59 flocks, respectively.